\documentclass{aa}
\usepackage[utf8]{inputenc}
\usepackage{amsmath}
\usepackage{amsfonts}
\usepackage{amssymb}
\usepackage{color,soul}
\usepackage{longtable}
\usepackage{graphicx, bm}

\begin{document}

\title{Constraining models of activity on comet 67P/Churyumov-Gerasimenko with Rosetta trajectory, rotation, and water production measurements}
\titlerunning{Activity models of comet 67P/Churyumov--Gerasimenko constrained by Rosetta}

\author{N.~Attree \inst{1, 2} \and L.~Jorda \inst{1} \and O.~Groussin \inst{1} \and S.~Mottola \inst{4}
\and N.~Thomas \inst{3} \and Y.~Brouet \inst{3}
\and E.~K{\"u}hrt \inst{4} \and  M.~Knapmeyer \inst{4} \and F.~Preusker \inst{4} \and F.~Scholten \inst{4} \and J.~Knollenberg \inst{4} \and S.~Hviid \inst{4}
\and P.~Hartogh \inst{5} \and R.~Rodrigo \inst{6}
}

\institute{Aix Marseille Univ, CNRS, CNES, Laboratoire d'Astrophysique de Marseille, Marseille, France
\and Earth and Planetary Observation Centre, Faculty of Natural Sciences, University of Stirling, UK \email{n.o.attree@stir.ac.uk}
\and Physikalisches Institut, Universit{\"a}t Bern, Sidlerstrasse 5, 3012 Berne, Switzerland
\and Deutsches Zentrum f{\"u}r Luft-und Raumfahrt (DLR), Institut f{\"u}r Planetenforschung, Rutherfordstra{\ss}e 2, 12489 Berlin, Germany
\and Max-Planck-Institut f{\"u}r Sonnensystemforschung, Justus-von-Liebig-Weg 3, 37077 G{\"o}ttingen, Germany
\and International Space Science Institute, Hallerstrasse 6, 3012 Bern, Switzerland
}

\abstract{}
   {We use four observational data sets, mainly from the Rosetta mission, to constrain the activity pattern of the nucleus of comet 67P/Churyumov-Gerasimenko.} 
   {We develop a numerical model that computes the production rate and non-gravitational acceleration of the nucleus of comet 67P as a function of time, taking into account its complex shape with a shape model reconstructed from OSIRIS imagery. We use this model to fit three observational data sets: the trajectory data from flight dynamics; the rotation state, as reconstructed from OSIRIS imagery; and the water production measurements from ROSINA, of 67P. The two key parameters of our model, adjusted to fit the three data sets all together, are the activity pattern and the momentum transfer efficiency (i.e., the so-called ``$\eta$ parameter'' of the non-gravitational forces).}
   {We find an activity pattern able to successfully reproduce the three data sets simultaneously. 
   The fitted activity pattern exhibits two main features: a higher effective active fraction in two southern super-regions ($\sim 10$~\%) outside perihelion compared to the northern ones ($< 4$~\%), and a drastic rise of the effective active fraction of the southern regions ($\sim 25-35$~\%) around perihelion. We interpret the time-varying southern effective active fraction by cyclic formation and removal of a dust mantle in these regions. Our analysis supports moderate values of the momentum transfer coefficient $\eta$ in the range $0.6-0.7$; values $\eta\leq0.5$ or $\eta\geq0.8$ degrade significantly the fit to the three data sets. Our conclusions reinforce the idea that seasonal effects linked to the orientation of the spin axis play a key role in the formation and evolution of dust mantles, and in turn largely control the temporal variations of the gas flux.}
   {}
   
\keywords{comets: general, comets: individual (Churyumov-Gerasimenko), planets and satellites: dynamical evolution and stability}

\maketitle

\section{Introduction}

The sublimation of ices when a comet is injected from its reservoir to the inner solar system triggers the emission of molecules.
This outgassing produces in turn a reaction force which can accelerate the comet nucleus in the opposite direction.
The perturbing effect of cometary activity on the trajectory of comets has been established in the 1950's in the pioneering work by \citet{Whipple50}.
At that time, it was clear that most comet trajectories were affected by a significant nongravitational acceleration (hereafter ``NGA'') linked to their activity around perihelion \citep{Marsden68}.
Shortly after, a theoretical model describing the nongravitational force (hereafter ``NGF'') produced by the sublimation of ices was established by \citet{Marsden1973}.
This model was based on a simple function describing the heliocentric dependence of the sublimation of water ice of a comet, combined with constant scaling parameters $A_1$, $A_2$ and $A_3$ describing the amplitude of the NGA along the three components (radial, transverse, normal) in the orbital frame of the comet.
The model has been modified by \citet{Yeomans89} to incorporate an asymmetric term used to describe the shift of the maximum of activity with respect to perihelion.
It is worth mentioning here that these simple models are still in use nowadays to fit astrometric measurements, and hence to describe cometary orbits.\\

More sophisticated models have been introduced since then.
Images acquired during the flyby of comet 1P/Halley by Giotto showed narrow dust jets \citep{Keller87}, leading to the idea that the activity may be confined to localised areas.
This led to a new model of nongravitational acceleration \citep{Sekanina93} in which the outgassing originates from discrete areas at the surface of a rotating nucleus.
\citet{Szutowicz00} used Sekanina's model to fit astrometric measurements of comet 43P/Wolf-Harrington obtained during nine perihelion passages.
She also compared the modelled production rate with visual lightcurves used as a proxy for the activity of the comet \citep{Szutowicz00}.
Recently, \citet{Maquet12} revisited Sekanina's approach with a model in which the activity is parametrised by the surface fraction of exposed water ice in ``latitudinal bands'' at the surface of an ellipsoidal nucleus.
More accurate ground-based measurements as well as space missions to comets offered the opportunity to incorporate new physical processes in models of the NGA.
\citet{Rickman89} used the change of the orbital period of comet 1P/Halley caused by the NGA to extract its mass and density.
The detailed description of the local outflow velocity incorporated a ``local momentum transfer coefficient'' $\zeta_l$, originally called $\eta$ in the improved description of the solid-gas interface introduced by \citet{Crifo1987}.
This coefficient represents the fraction of the emitted gas's momentum (dependent on its thermal velocity) which needs to be considered in the calculation of the momentum transfer, and thus of the NGF \citep[see][]{Crifo1987}.
\citet{Davidsson04} used 2D thermal modelling including thermal inertia, self-shadowing, self-heating and an activity pattern to fit the NGA of comet 19P/Borrelly.
They could retrieve the mass of the nucleus and constrain the direction of the spin axis.
The method was also applied to comets 67P/Churyumov-Gerasimenko~\citep{Davidsson05}, 81P/Wild~2 \citep{Davidsson06}, and 9P/Tempel~1 \citep{Davidsson07}, all targets of space missions.\\

The exploration of comet 1P/Halley in 1986 also yielded the discovery of the non-principal axis rotation of this comet \citep{Samarasinha91}.
Since then, the torque of the NGF was thus identified as the major effect responsible for changes of the rotational parameters \citep[and references therein]{Samarasinha04}.
Its modelling is required to understand the observed change in the rotational parameters of cometary nuclei, as well as the apparition of non-principal axis rotations.
Changes in the spin period of several comets have been detected from the analysis of lightcurves \citep{Mueller96,Gutierrez03,Samarasinha2004,Drahus06,Knight11, Bodewits2018}.
Predictions of the expected change in the direction of the spin axis and spin period for comets 81P/Wild~2 \citep{Gutierrez07} and 67P/Churyumov-Gerasimenko \citep[hereafter ``67P'',][]{Gutierrez05} have been made.
Recently, a change in the spin period of comet 67P has been clearly identified between the 2009 and 2015 perihelion passages \citep{Mottola14} based on early images acquired by the OSIRIS camera aboard Rosetta.
\citet{Keller} showed that the spin period variation curve of 67P is controlled at first order by the bilobate shape of the nucleus.

Thanks to its long journey accompanying comet 67P, Rosetta provides a unique chance to record measurements of most parameters involved in the modelling of the NGF.
The mass of the comet has been retrieved by \citet{Patzold16} from the radio science experiment aboard the spacecraft.
The shape has been retrieved from a stereophotogrammetric analysis of a subset of OSIRIS images \citep{Preusker17}, leading to an accurate knowledge of the moments of inertia.
The activity pattern has been constrained in the early phase of the mission by \citet{Marschall16,Marschall17} from ROSINA measurements in the form of ``effective active fractions'' associated to geological regions.
The total water production rate has been constrained from ROSINA measurements, complemented by ground-based observations \citep{Hansen}.
Finally, the spin period has been monitored throughout the mission by ESA's flight dynamics and OSIRIS teams before \citep{Jorda2016} and after \citep{Kramer2019} perihelion.
The latter \citep{Kramer2019} modelled the temporal evolution of the rotational parameters, comparing it with the measured change in spin period and spin axis orientation.
The aim of this article is to try to reproduce three data sets derived by several Rosetta instruments (see section~\ref{observations}) with a model of the NGF (see section~\ref{modelling}) in order to retrieve: {\it (i)} the local effective active fraction and its temporal variations around perihelion, and {\it (ii)} a recommended value for the momentum transfer coefficient $\eta$ (see section~\ref{results}).
The results will be discussed in section~\ref{discussion}, together with recommendations for the calculation of NGF of other comets.

\section{Observational constraints}
\label{observations}

In this section, we describe the observational data, mainly obtained by the Rosetta spacecraft, to which we will attempt to fit our NGF model.

\subsection{Water production rate}

The total water production rate is an important constraint for any model of cometary activity, and a significant effort has been made to measure it for 67P. 
As summarised by \cite{Hansen}, a number of different instruments have all been used and comparing and synthesising their results in non-trivial. 
Here, we use the ROSINA (Rosetta Orbiter Spectrometer for Ion and Neutral Analysis) measurements, empirically corrected for spacecraft position, as our observed data points, $O_{Q}$.
\cite{Hansen} describe how estimating the uncertainty in these data, $\sigma_{Q}$, is difficult so we use the bounds on the power-law, fitted with heliocentric distance to the inbound ROSINA data, as given in their Table~2.

We point out that ROSINA data are inferred from local measurements in the coma of 67P.
Around perihelion, the spacecraft was located at a distance of $200\ \mathrm{km}$ from the nucleus, making it difficult to infer a total production rate, whilst ground-based observations (see, for example, \citealp{Bertaux}) have also suggested a variation in peak production between perihelion passages. Production rate estimates from Rosetta's line-of-sight instruments MIRO and VIRTIS are also generally lower that the \citet{Hansen} results. These are important caveats to bear in mind when interpreting our results.
Finally, the possibility that sublimating icy grains are emitted by the nucleus is not considered in this article, and neither are other gas species, such as CO$_{2}$, CO and O$_{2}$. Other species represent $<10\%$ of the gas number density at perihelion \citep{Hansen} and their production curves are not as well constrained as water. Therefore, for this study we choose to focus on water, which is the primary driver of non-gravitational forces.

\subsection{Trajectory}
\label{trajectory_data}

Outgassing produces a back-reaction force, and a resulting non-gravitational acceleration (NGA), on cometary nuclei which affects their trajectory in a measurable way. 
For 67P, the nucleus trajectory has been reconstructed by the flight dynamics team of ESA, by a combination of radio-tracking of the Rosetta spacecraft from Earth and optical navigation of the comet relative to it, and is available in the form of NASA SPICE kernels \citep{Acton1996}.
Therefore, the 3D position of the comet in a heliocentric reference frame has an accuracy greatest in the Earth-comet range direction ($\bm{R}$, claimed accuracy of $\sim 10\ \mathrm{m}$) and much less in the perpendicular (cross-track) directions (claimed accuracy of $\sim 100\ \mathrm{km}$).

\begin{figure}
\sidecaption
\includegraphics[width=9.5cm]{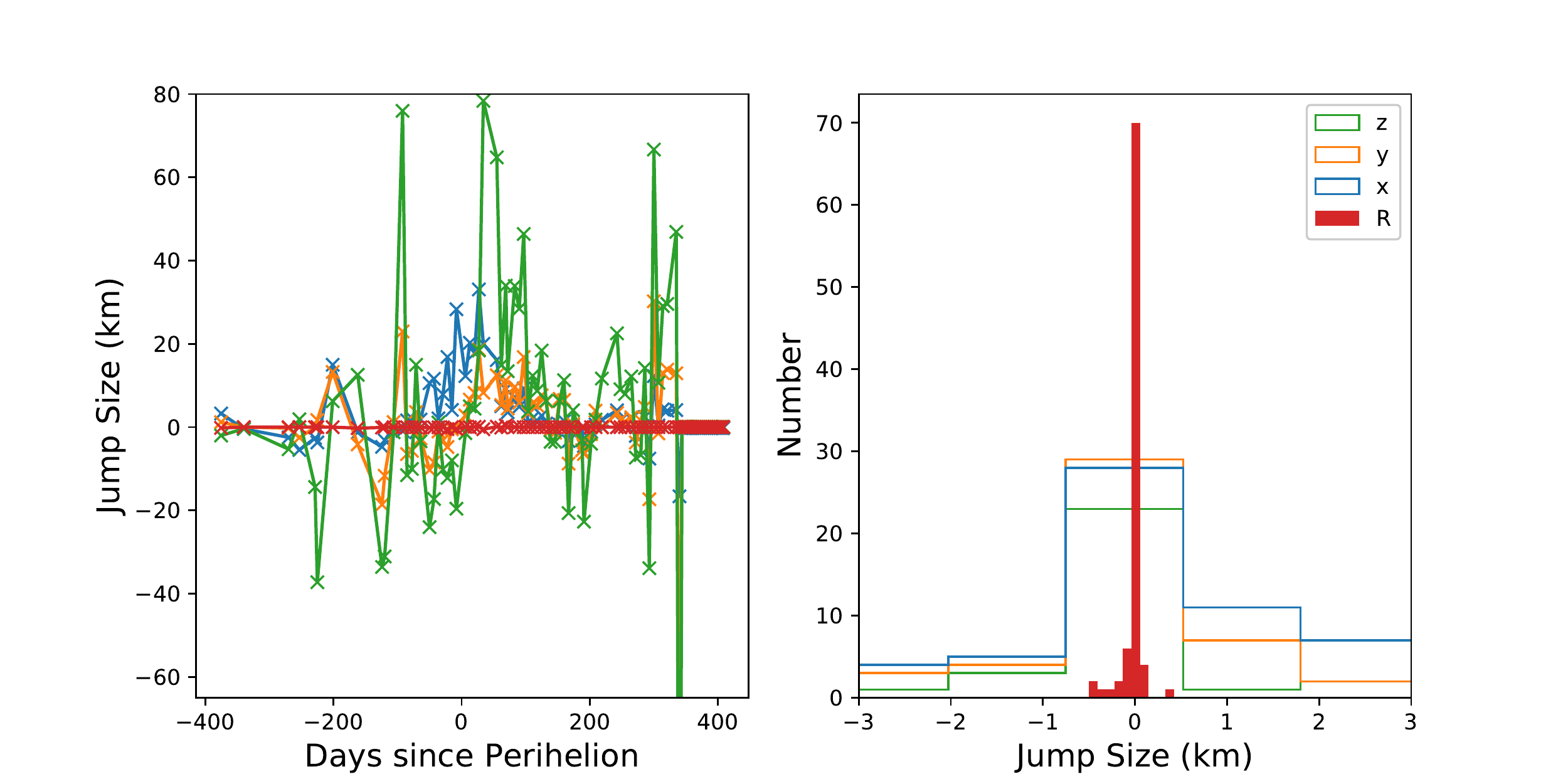}
\caption{Discontinuities identified in the position of comet 67P from the SPICE kernels, in $(x, y, z)$ heliocentric J2000 coordinates and Earth-comet range ($R$). On the left: as a function of time and on the right as a histogram of sizes.}
\label{jumpsizes}
\end{figure}

Theoretically, the NGA resulting from outgassing could be directly extracted from the residuals between this measured trajectory and a modelled gravitational orbit (taking into account general relativity and the gravitational accelerations of all major planets, Pluto and the most massive asteroids). 
During the course of this work, however, it was discovered (and later confirmed by ESAC; B.~Grieger, personal communication) that the reconstructed comet and spacecraft trajectories contain a series of discontinuities, at which the objects' positions vary over hundreds of metres to several kilometres in an instantaneous time, making the above method impossible. 
Within the orbital segments between these `jumps', there is no difference, to machine precision, between our own orbital integrations (see Sect.~\ref{NGA_model} below) and the reconstructed trajectory, demonstrating that it is a purely gravitational solution. 
The jumps occur at the boundaries between the integration segments that make up the reconstructed orbit, and represent the offset in the objects' positions over the course of each segment due to the un-modelled non-gravitational acceleration. 
Unfortunately, it proved impossible to extract the NGAs directly from the jumps themselves as they contain, not only, the NGA effect but also the typical uncertainty in the state vector at the start of each segment, which is of a comparable magnitude. 
The jumps therefore have random magnitudes with time (Fig.~\ref{jumpsizes}) and must be considered an additional source of noise in the uncertainty in the measured positions.

Despite these issues, the reconstructed kernels remain a good description of the comet's trajectory over orbital timescales, and within the limits of accuracy of the typical jump size. 
Therefore it is still possible to compare these measurements with a model of the orbit, including a thermal outgassing NGA as well as N-body gravitational interactions, in order to constrain said model. 
To do so, we use the magnitude of the comet-to-Earth-centre range as our observable $O_{R}$, since the jump sizes are smallest in this direction (Fig.~\ref{jumpsizes}). 
Considering the jumps to be a source of random error, we conservatively estimate the uncertainty in $O_{R}$ as $\sigma_{R} \approx 1$ km.

%
%
%

\subsection{Rotation}

Back-reaction from outgassing not only produces a net acceleration on the nucleus but can also, depending upon its shape, produce a net torque, altering its rotation state. 
The rotational parameters of 67P, including its spin rate over time, $\omega(t)$, has been measured as part of the reconstruction of its 3D shape from OSIRIS images \citep{Jorda2016}.
%
%

\begin{figure}
\resizebox{\hsize}{!}{\includegraphics{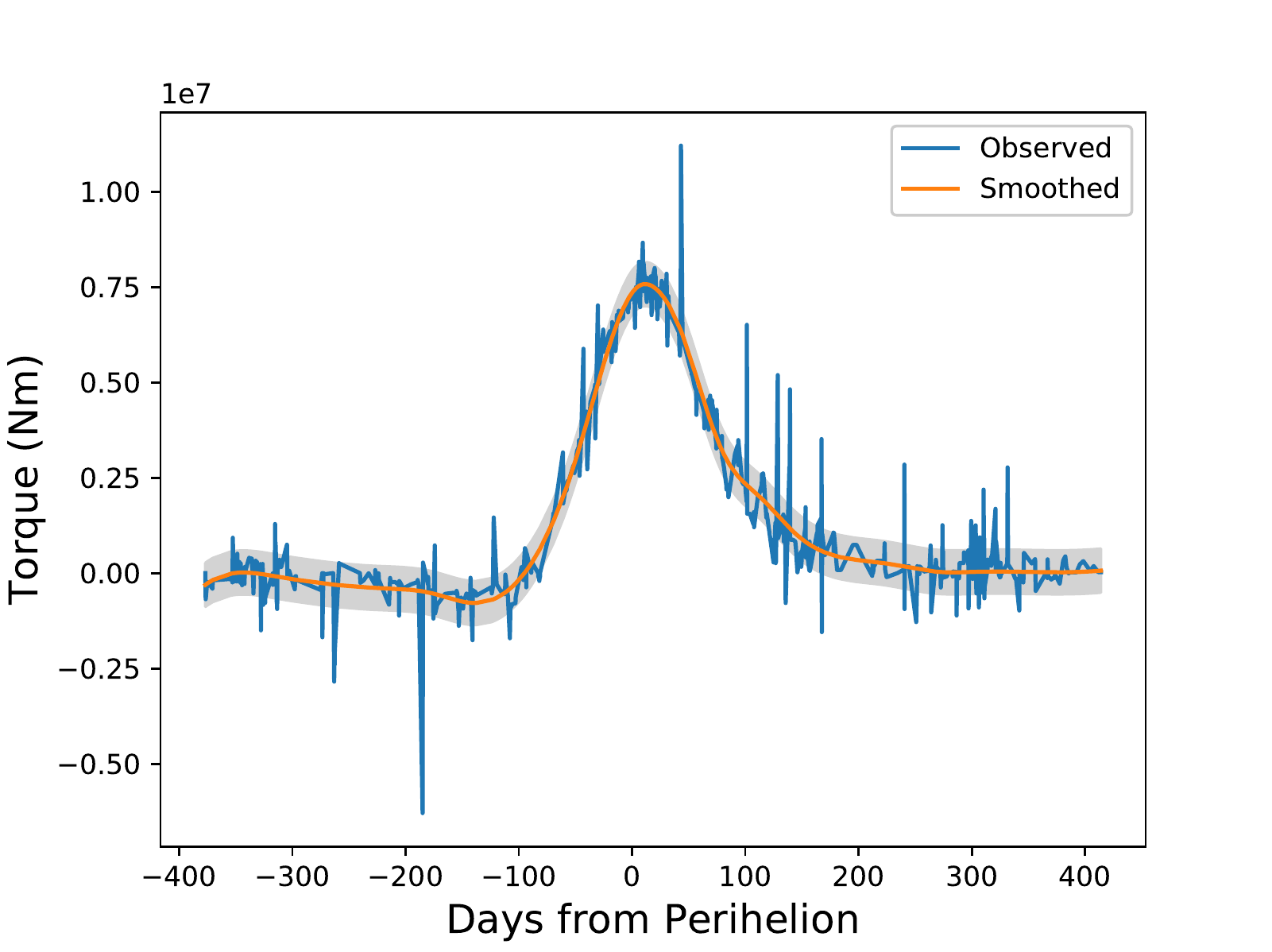}}
\caption{Observed torque, derived from the 67P rotation state from OSIRIS measurements, and a smoothed cubic spline fit to the data. The grey region represents the RMS of the residuals between the two.}
\label{torque_data}
\end{figure}

After verifying that the cross terms relating to the angular velocities along the first and second principal axes of the comet are negligible, changes in spin rate, $\dot{\omega_{z}}$, can be directly related to the $z$ component of the torque ($\tau_{z}$, in a body-fixed frame) by Eq.(\ref{torque}), where $I_{z}=1.899\times10^{19}$ kg m$^{2}$ is the third (largest) moment of inertia derived from the shape model assuming a constant density of $538\ \mathrm{kg}\ \mathrm{m}^{-3}$ \citep{Preusker17}. 
\begin{equation}
\tau_{z} \approx I_{z} \dot{\omega_{z}},
\label{torque}
\end{equation}
Differentiating the observed $\omega(t)$ by time exacerbates measurement uncertainties so that the produced torque curve becomes extremely noisy. 
For comparing with our simulations below, we therefore smooth the data by fitting it to a cubic spline, as shown in Fig.~\ref{torque_data}. 
Our fitted spline is then used as the torque observable, $O_{\tau}$, with an assumed uncertainty equal to the root mean squared residuals of the derived minus smoothed data, $\sigma_{\tau}=575000\ \mathrm{N}.\mathrm{m}$ (shown in grey bounds in Fig.~\ref{torque_data}).

\section{Modelling}
\label{modelling}

\subsection{Thermal model}
A thermal model is required to compute the temperature of the sublimating layer (assumed to be at the surface on the nucleus), from which we can derive the non-gravitational forces acting on the nucleus. Our thermal model takes into account solar insolation, surface thermal emission, sublimation of water ice, projected shadows and self-heating. We use a decimated version of the 67P shape model called {\it SHAP7} \citep{Preusker17}, with $124\ 938$ facets. The temperature is computed for each facet of the shape model, 36 times per rotation (i.e., every 1240~s), and 70 times (i.e., every $\sim 10$~days) over the 2~years of the Rosetta mission, to ensure a good temporal coverage. At each time step, the distance to the Sun, the orientation of each facet relative to the Sun, and the projected shadows, are computed using the OASIS software \citep{Jorda2010}. Due to the large number of facets (>100 000) and time steps (>2500), heat conduction is neglected in the thermal model for numerical reasons. To test this assumption, we computed the production rate (Eq.~\ref{z}) and acceleration (Eq.~\ref{NGF}) of a spherical nucleus, at perihelion (where the torque is maximum), for two cases: a thermal inertia of 0 and 100~J/m$^2$/K/s$^{0.5}$. The production rate and the acceleration are $\sim$7~\% smaller for a thermal inertia of 100~J/m$^2$/K/s$^{0.5}$ (compared to a null thermal inertia), and the direction of the acceleration vector differs by less than 3~deg. Neglecting the thermal inertia therefore appears reasonable compared to the data uncertainties (e.g., production rates).

The surface energy balance of the thermal model is given by Eq.~(\ref{thermal_model}), for a facet with index $i$, where $A_b=0.0119$ is the Bond albedo at 480 nm \citep{Fornasier15}, $F_{\rm sun}=1370$ W/m$^2$ is the solar constant, $z_i$ is the zenithal angle, $r_h$ the heliocentric distance, $SH_i$ is the self-heating given by Eq.~(\ref{self_heating}), $\epsilon=0.95$ is the assumed infrared emissivity, $T_i$ is the surface temperature, $f_i$ is the fraction of water ice (in our case, either 0 for pure dust or 1 for pure ice, see below), $\alpha=0.25$ accounts for the recondensation of water ice on the surface \citep{Crifo1987}, $L=2.66\times10^6$ J/Kg is the latent heat of sublimation of water ice at 200 K, and $Z_i$ is the water sublimation rate given by Eq.~(\ref{z}). 
\begin{equation}
\frac{(1-A_b)F_{\rm sun}\cos z_i}{r_{\rm h}^2} +SH_i= \epsilon \sigma T_i^4 + f_i(1-\alpha)LZ_i(T_i) 
\label{thermal_model}
\end{equation}In Eq.~(\ref{self_heating}), the self-heating $SH_i$ is the sum of the infrared flux coming from all the $n$ facets with index $j$ that see facet $i$, where $S_j$ is the surface of facet $j$, $\theta_j$ the angle between the normal to facet $j$ and the vector joining facets $i$ and $j$,  $\theta_i$ the angle between the normal to facet $i$ and the vector joining facets $i$ and $j$, and $d_{ij}^2$ is the distance between facets $i$ and $j$. This formalism is similar to that of \citet{Guitierez2001}. To look for which facets are seeing each others, we used an algorithm developed at Laboratoire d'Astrophysique de Marseille based on ray tracing and hierarchical search. The computation of the viewing factors $(S_j \cos \theta_j \cos \theta_i)/(\pi d_{ij}^2)$ is purely geometric and depends only on the shape model: it is therefore only performed once at the beginning.\\
\begin{equation}
SH_i = \sum_{j=0}^{n} \epsilon \sigma T_j^4\frac{S_j \cos \theta_j \cos \theta_i}{\pi d_{ij}^2}
\label{self_heating}
\end{equation}In Eq.~(\ref{z}), the sublimation mass flow rate per facet is calculated with the molar mass $M = 0.018$ kg, the two constants $A=3.56\times10^{12}$ Pa and $B = 6162$ K for water \citep{FanaleSalvail}, and the gas constant $R = 8.3144598$ J K$^{-1}$ mol$^{-1}$.
\begin{equation}
Z_i = A e^{-B/T_{ice}} \sqrt{\frac{M}{2\pi R T_{ice}}}
\label{z}
\end{equation}
Finally, to compute the non-gravitational forces (Sect. 3.2), we need the sublimation rate (Eq.~(\ref{z})) and the gas velocity (Eq.~(\ref{v})), which both depend on a different temperature: the temperature of water ice $T_{\rm ice}$ for $Z_i$, and the temperature of dust $T_{\rm dust}$ for $v_i$. We therefore run our thermal model (Eqs.~(\ref{thermal_model}) to (\ref{z})) with two extreme cases: [1] with $f_i=0$, which corresponds to a pure dust model, to compute $T_{\rm dust}$, and [2] with $f_i=1$, which corresponds to a pure water ice model, to compute $T_{\rm ice}$.

\subsection{Non-Gravitational Force model}
\label{NGA_model}

The reaction force vector per facet is then calculated based on this mass flow rate, and the total acceleration is the sum over all facets divided by the comet mass ($M_{67P} = 9.982\times10^{12}$ kg; \citealp{Patzold16}):
\begin{equation}
\bm{F_{i}} = - \eta x_{i} Z_{i} v_{i} \bm{S_{i}},\\
\bm{a}_{NG} = \frac{\sum_{i}{\bm{F_{i}}}}{M_{67P}}.
\label{NGF}
\end{equation}

where $\eta$ is the momentum transfer coefficient \citep{Crifo1987,Rickman89}, which we here assume to be constant across the comet. $\bf{S_{i}}$ is the surface area of each facet (in the direction of its normal) and $x_{i}$ is its effective active fraction. Mass flow rate is calculated with Eq.~(\ref{z}) and the gas velocity is taken as the thermal velocity
\begin{equation}
v_{i} = \sqrt{\frac{8RT_{dust}}{\pi M}},
\label{v}
\end{equation}
assuming equilibrium with the surface grey body temperature, i.e.~that of the dust from run [1] of the thermal model. This is the upper limit that the gas can reasonably reach, meaning our non-gravitational force will be on the high end of estimation and our effective active fractions are lower limits. Calculated dust temperatures range between $\sim20-390$ K (ice temperatures are limited by the sublimation to $\sim200$ K), leading to thermal velocities of $\sim155-658$ m s$^{-1}$.

Water production and torque are likewise summed over the surface

\begin{equation}
Q = \sum_{i}{x_{i}Z_{i}S_{i}}.\\
\bm{\tau}_{NG} = \sum_{i}{\bm{\tau_{i}}},
\label{Q}
\end{equation}

where torque per facet is the vector product of each force vector with its radius vector to the centre of mass ($\bm{r_{i}}$). This can also be expressed as the magnitude of the force multiplied by a ``torque efficiency'' (see \citealp{Keller}):
\begin{equation}
\bm{\tau}_{i} = \bm{r_{i}} \times \bm{F_{i}} = F_{i} (\bm{r_{i}} \times \bm{\hat{S_{i}}}).
\label{NGtorque}
\end{equation}

\begin{figure}
\resizebox{\hsize}{!}{\includegraphics{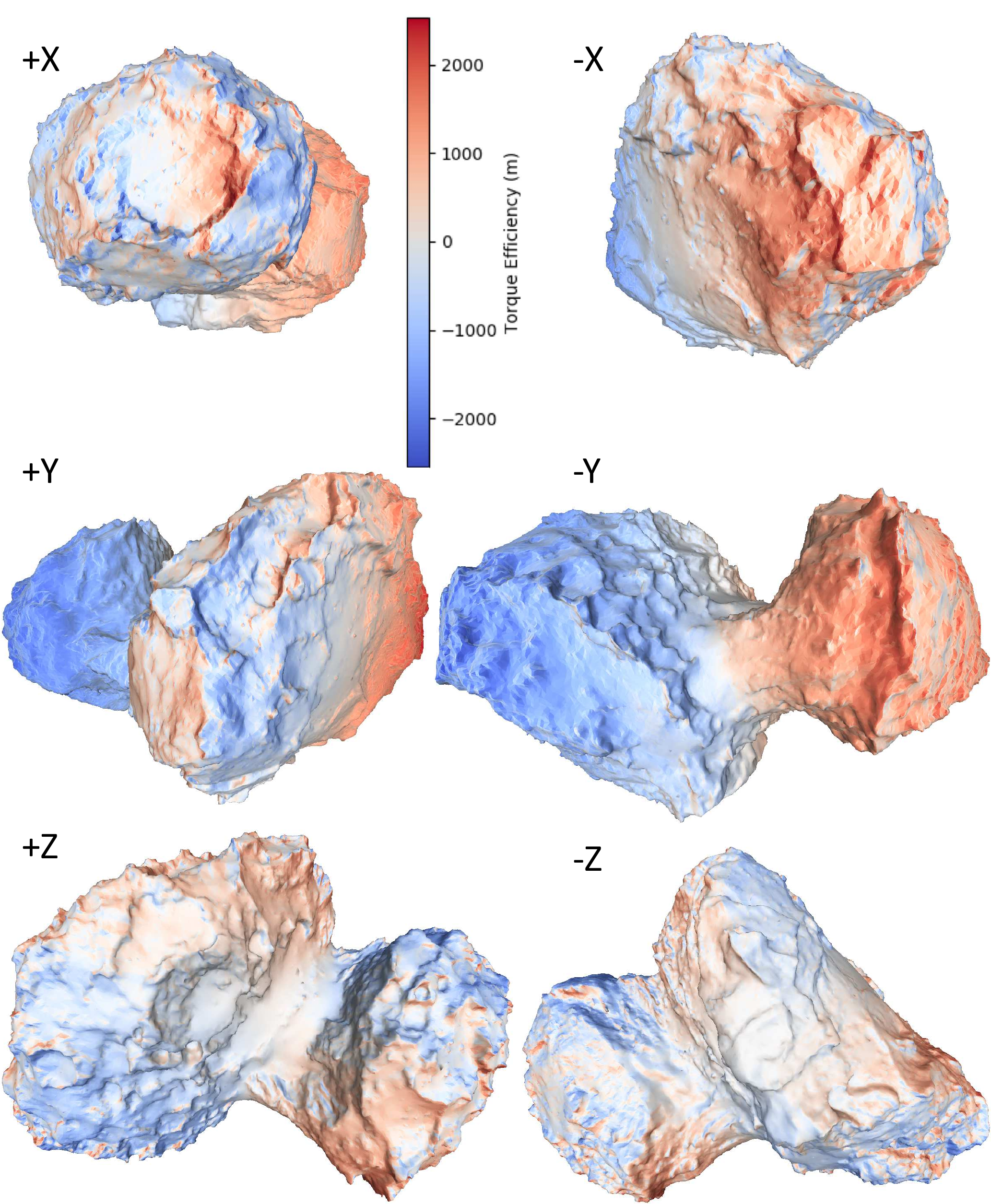}}
\caption{Torque efficiency in metres, a factor determined by the geometry as defined in Eqn.~\ref{NGtorque}. The comet's rotation axis is in the $+z$ direction through the `neck' region in the centre.}
\label{torqueefficiency}
\end{figure}

The $z$ component of the total net torque can then compared with the observations, using Eq.~(\ref{torque}). It is advantageous to use the torque efficiency formalism since this vector is in the body-fixed frame, and can be calculated once at the beginning of the simulation run, rather than being recalculated each time. Torque efficiency is also a useful way of visualising the effects of differing spatial distributions in activity, which will be important later during the optimisation. Mapping torque efficiency onto the shape model (Fig.~\ref{torqueefficiency}) shows how local variations in topography combine with large scale orientations of regions, varying the effects of activity on the comet's rotation across its surface (compare with Fig.~1 in \citealp{Keller}, which uses an older, incomplete shape model that is missing the southern hemisphere).

Non-gravitational forces are calculated for each of the $70\times36=2520$ runs of the thermal model and the relevant quantities (force and torque vectors and summed water production) are averaged over a day. This produces smoothly time-varying curves which can be inspected at any time of interest, using bilinear interpolation, which we refer to as our model solution.
For comparison with the observed water production rate and torque, we can simply evaluate our model at the time of each measurement, producing $C_{Q}$ and $C_{\tau}$, and directly compare.

For the trajectory, however, a full N-body integration must be performed and the resulting position compared at each time ($C_{R}$). To do this we use the open-source $REBOUND$ code\footnote{\url{http://rebound.readthedocs.io/en/latest/}} \citep{ReinLiu}, complete with full general relativistic corrections \citep{Newhall} as implemented by the $REBOUNDx$ extension package\footnote{\url{http://reboundx.readthedocs.io/en/latest/index.html}}. All the major planets are included, as well as Ceres, Pallas, and Vesta, and are initialised with their positions in the J2000 ecliptic coordinate system according to the same ephemerides used in the Rosetta trajectory reconstructions (NASA/JPL solar system solution DE405; \citealp{Standish}). An additional particle representing 67P is initialised with its position given by SPICE. The system is then integrated forward in time, using the IAS15 integrator \citep{ReinSpiegel} and the standard equations of motion, with the addition of a custom acceleration, $\bm{a}_{NG}$, for 67P, provided by our model. The modelled comet begins to diverge from the measured positions and at each time of interest we directly compare the magnitudes of the computed and measured comet-to-Earth-centre ranges, $R$ (the most accurate part of the trajectory as described in sect.~\ref{trajectory_data} above).

\subsection{Optimisation}

In order to constrain the unknown parameters in our model, such as the the surface active fraction, we perform a bounded least-squares fit to the data using the dogleg optimisation routine \citep{Voglis} provided in the scientific Python package. The optimisation proceeds, attempting to minimise the standard objective function
\begin{equation}
Obj = \sum_{j=0}^{N}{\left(\frac{O_{j}-C_{j}}{\sigma_{j}}\right)^{2}},
\label{objective}
\end{equation}
with observed minus computed residuals, $O-C$, and observation uncertainties, $\sigma$, at each time-step, $j$, up to the total $N$. The root mean squared residuals are then $RMS = \sqrt{Obj/N}$.

In our case we have three separate datasets to fit to ($R$, $Q$ and $\tau$), a multi-objective optimisation problem, and we therefore use a linearly-scaled combination of the three to generate a combined objective function. The term inside the brackets in Eq.~(\ref{objective}) then becomes the sum of the three terms
\begin{align*}
\lambda_{R} {\left(\frac{O_{Rj}-C_{Rj}}{\sigma_{Rj}}\right)}, & \quad \text{for} \quad 0 < j \le N_{R},\\
\lambda_{Q} {\left(\frac{O_{Qj}-C_{Qj}}{\sigma_{Qj}}\right)}, & \quad \text{for} \quad N_{R} < j \le N_{R} + N_{Q}, \\
\lambda_{\tau} {\left(\frac{O_{\tau j}-C_{\tau j}}{\sigma_{\tau j}}\right)}, & \quad \text{for} \quad N_{R} + N_{Q} < j \le N, \tag{10}
\end{align*}
\label{combined_objective}

respectively, where $N = N_{R} + N_{Q} + N_{\tau}$ runs over the combined number of points in all three datasets, and the $\lambda$ scaling coefficients are variables, which themselves must be optimised in order to give the desired weighting to each dataset. We set $\lambda_{R} = 1$ and scale the other lambdas relative to it, by bootstrapping from the residuals of a preliminary run, to give roughly equal weighting to all three datasets. Due to the very small relative errors for range $(\sigma_{R}/R\sim\pm 1$ km $/1$ AU), this results in large values for the other lambdas ($\lambda_{Q}\sim\lambda_{\tau}\sim50$) to give equal weighting.

For each optimisation, we fix the coupling factor, $\eta$, at a constant value and parametrise the model in terms of effective active fraction, $x$, across the surface. The effects of $\eta$ on the best-fit model can then be studied independently. To begin with, we use the division of 67P's surface into five `super regions', as performed by \citet{Marschall16, Marschall17} in fitting ROSINA/COPS and OSIRIS data, and start the optimisation with initial values from their results. This divides the comet into: a southern hemisphere region, an equatorial region, a region covering the base of the body and top of the head, Hathor, and Hapi, as shown in Fig.~\ref{Map_5regions} below. The fitting routine then proceeds to optimise these five free parameters, subject to the lower and upper boundaries of zero and one, i.e.~$0-100\%$ active fraction. As described below, we also use more detailed parameterisations, including all 26 regions of \cite{Thomas15} for 26 free parameters.

\begin{figure}
\resizebox{\hsize}{!}{\includegraphics{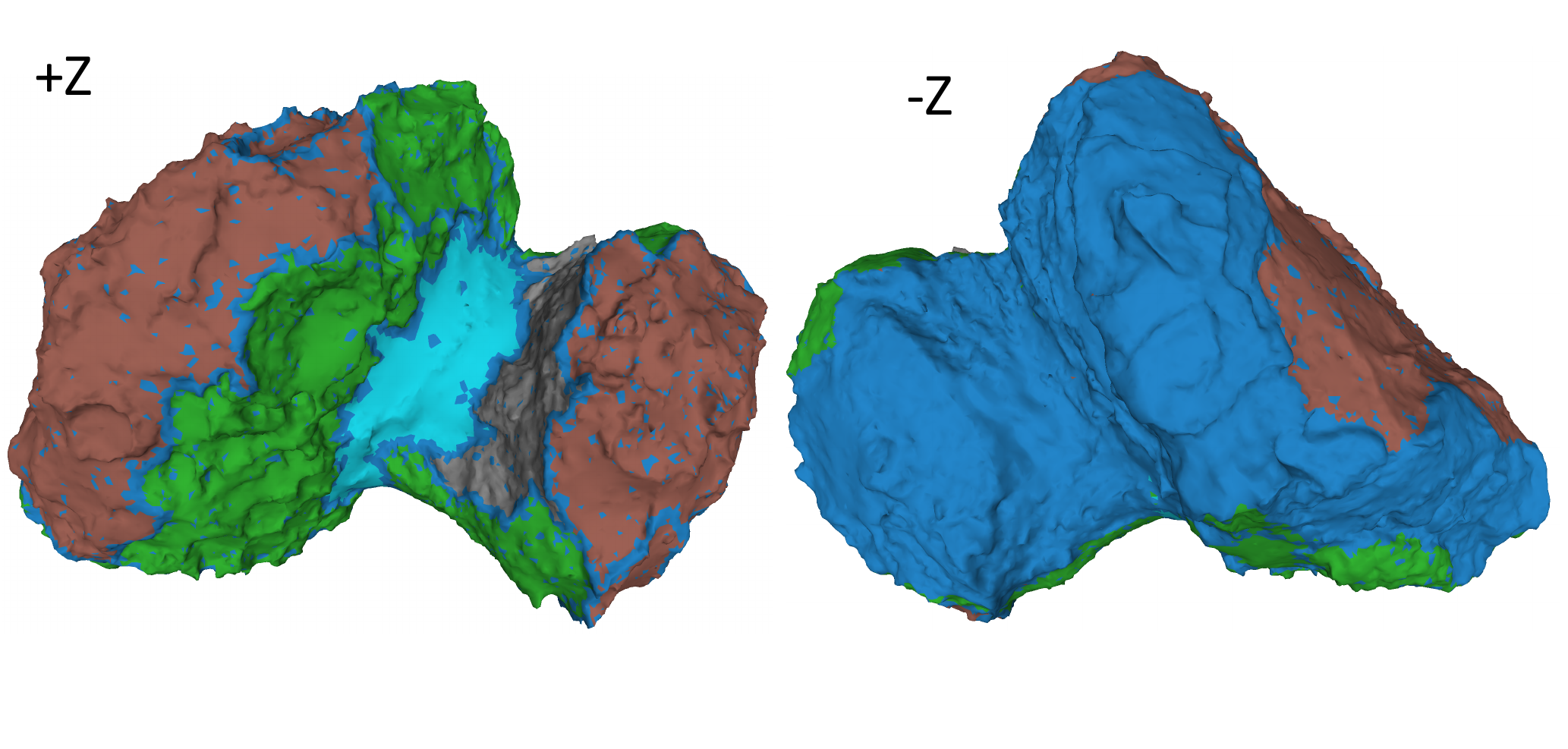}}
\caption{Map of the 5 `super regions' defined by \citet{Marschall16, Marschall17} and used in our solution A optimisation.}
\label{Map_5regions}
\end{figure}

\section{Results}
\label{results}

Before optimising our activity model we first test the N-body component by calculating the comet trajectory over the course of the Rosetta mission with no NGAs applied, and with the classic NGA parametrisation based on the model of \cite{Marsden1973}. This model computes the three components of $\bm{a}_{NG}$ (radial, along-orbit and normal-to-orbit) based on a power-law with heliocentric distance, and three scaling parameters $A_{1,2,3}$ found by a formal best-fit to the orbit for each comet. We use the $A_{1,2,3}$ values for the 2010 apparition of 67P from ground-observations given by NASA/JPL Horizons ephemerides\footnote{\url{https://ssd.jpl.nasa.gov/?horizons}}.

\begin{figure}
    \resizebox{\hsize}{!}{\includegraphics{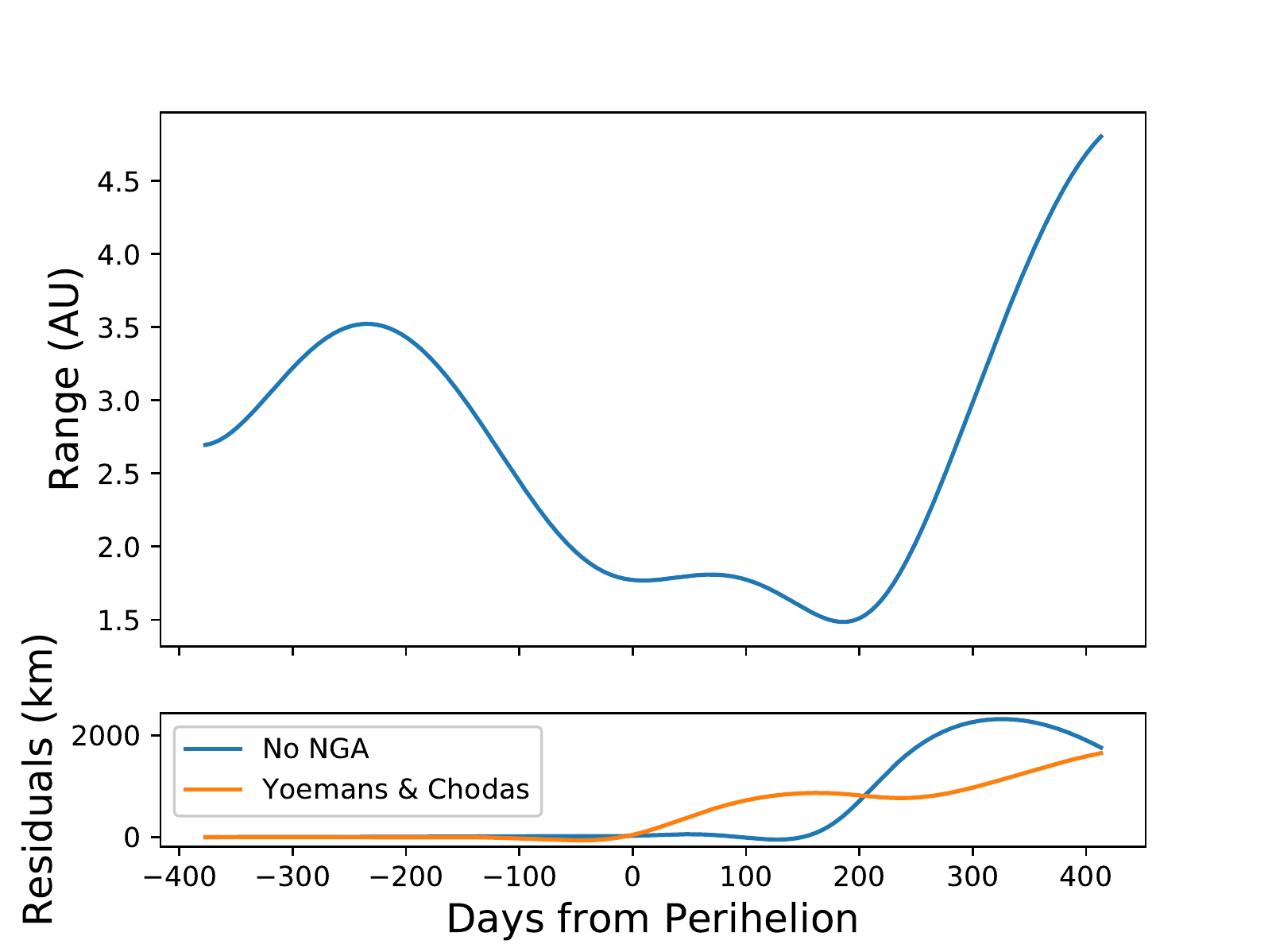}}
\caption{Observed Earth-comet range, $R$, and residuals for two models: a purely gravitational N-body solution with no NGA, and a ground-based solution based on the model of \cite{Marsden1973} (see text for details). Differences between the observed and computed solutions, as well as the jumps described in Sect~\ref{trajectory_data}, are invisible at the scale of the top plot.}
\label{base_models}
\end{figure}

Figure \ref{base_models} shows a plot of the observed range plus the residuals to both models. The RMS residuals are 1029~km and 675~km, respectively, showing the order of magnitude of the accuracy of the \cite{Marsden1973} model with ground-based observations of roughly arcsecond accuracy. This stands as a baseline, against which we can check our own activity model. 

\subsection{5 parameter solution (A)}

\begin{table*}
\begin{center}
\begin{tabular}{cccccrccr}
Solution & $\eta$ & $\lambda_{R}$ & $\lambda_{Q}$ & $\lambda_{\tau}$ & RMS$_{R}$ (km) & RMS$_{Q}$ & RMS$_{\tau}$ & RMS$_{Obj}$ \\
\hline
No NGA       &     &   &    &    & 1029 & & & 1029 \\
Ground-based &     &   &    &    & 675 & & & 675 \\
A            & 0.7 & 1 & 50 & 50 & 163 & 6.6 & 5.1 & 259 \\ 
B            & 0.7 & 1 & 50 & 50 & 187 & 5.2 & 2.1 &  235 \\ 
{\bf C}      & {\bf 0.7} & {\bf 1} & {\bf 50} & {\bf 50} &  {\bf 46} & {\bf 4.2} & {\bf 0.57} & {\bf 125}\\ 
Ca           & 0.5 & 1 & 50 & 50 & 115 & 5.5 & 1.32 & 176 \\
Cb           & 0.6 & 1 & 50 & 50 &  64 & 4.3 & 0.61 & 130 \\
Cc           & 0.8 & 1 & 50 & 50 &  62 & 5.5 & 1.52 & 168 \\
\end{tabular}
\caption{Parameters and root-mean-squared residuals (in range, water production rate, non-gravitational torque, and the total objective function) for the best-fit models. `No NGA' is an N-body only model computed in REBOUND, while `Ground-based' has an additional force given by the \citet{Marsden1973} model with parameters from NASA Horizons. A, B and C use the thermal model described here. The best model (C) is highlighted in bold font.}
\label{table}
\end{center}
\end{table*}

\begin{figure}
\resizebox{\hsize}{!}{\includegraphics{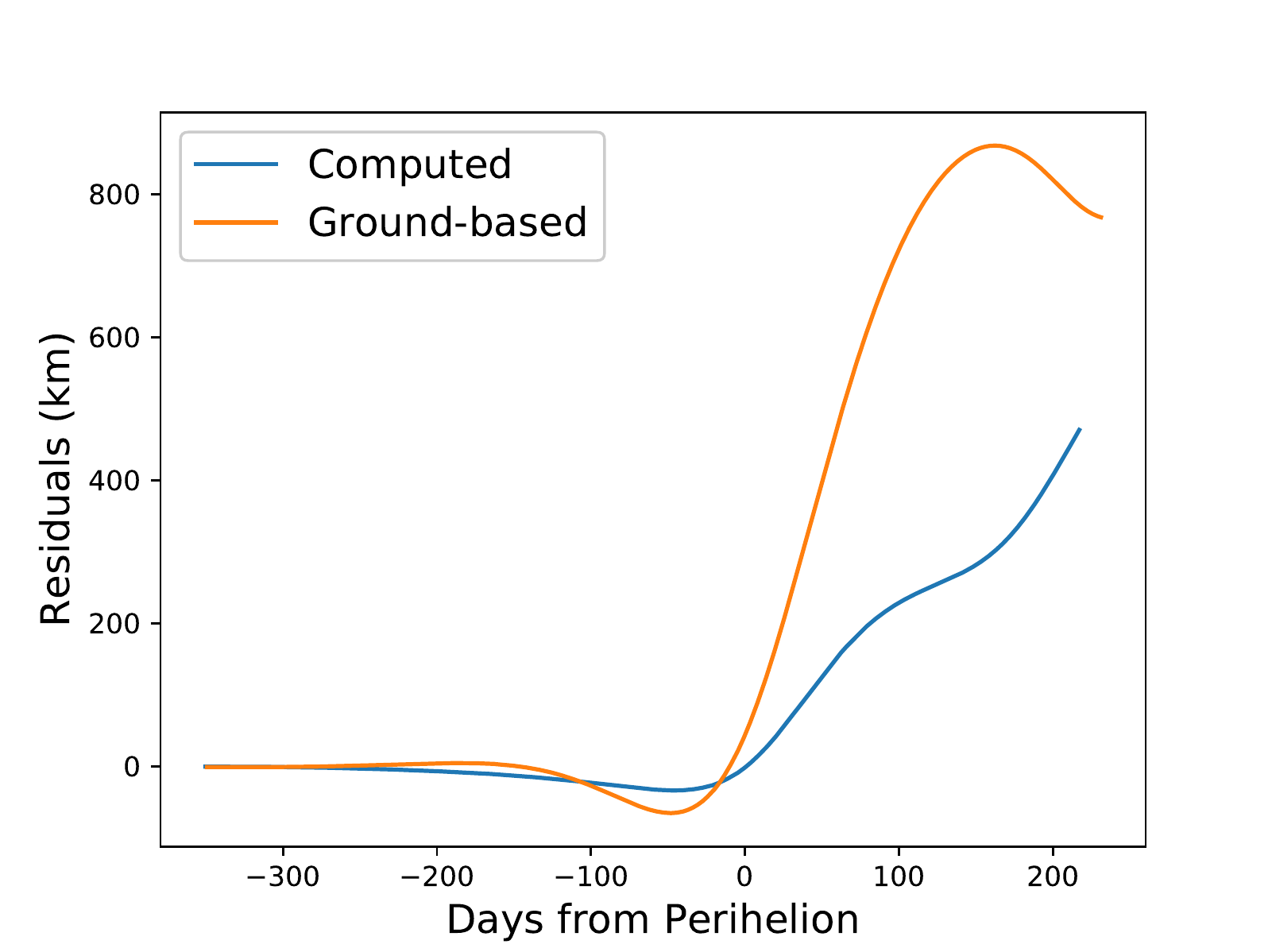}}
\caption{Observed minus computed range, $R$, for solution A (blue curve). For comparison, the residuals to the ground-based solution, using the \citet{Marsden1973} model as in Fig.~\ref{base_models}, are shown in orange. Both curves diverge from zero most sharply following the maximum perturbation around perihelion, but our solution is an improvement.}
\label{Range50}
\end{figure}

\begin{figure}
\resizebox{\hsize}{!}{\includegraphics{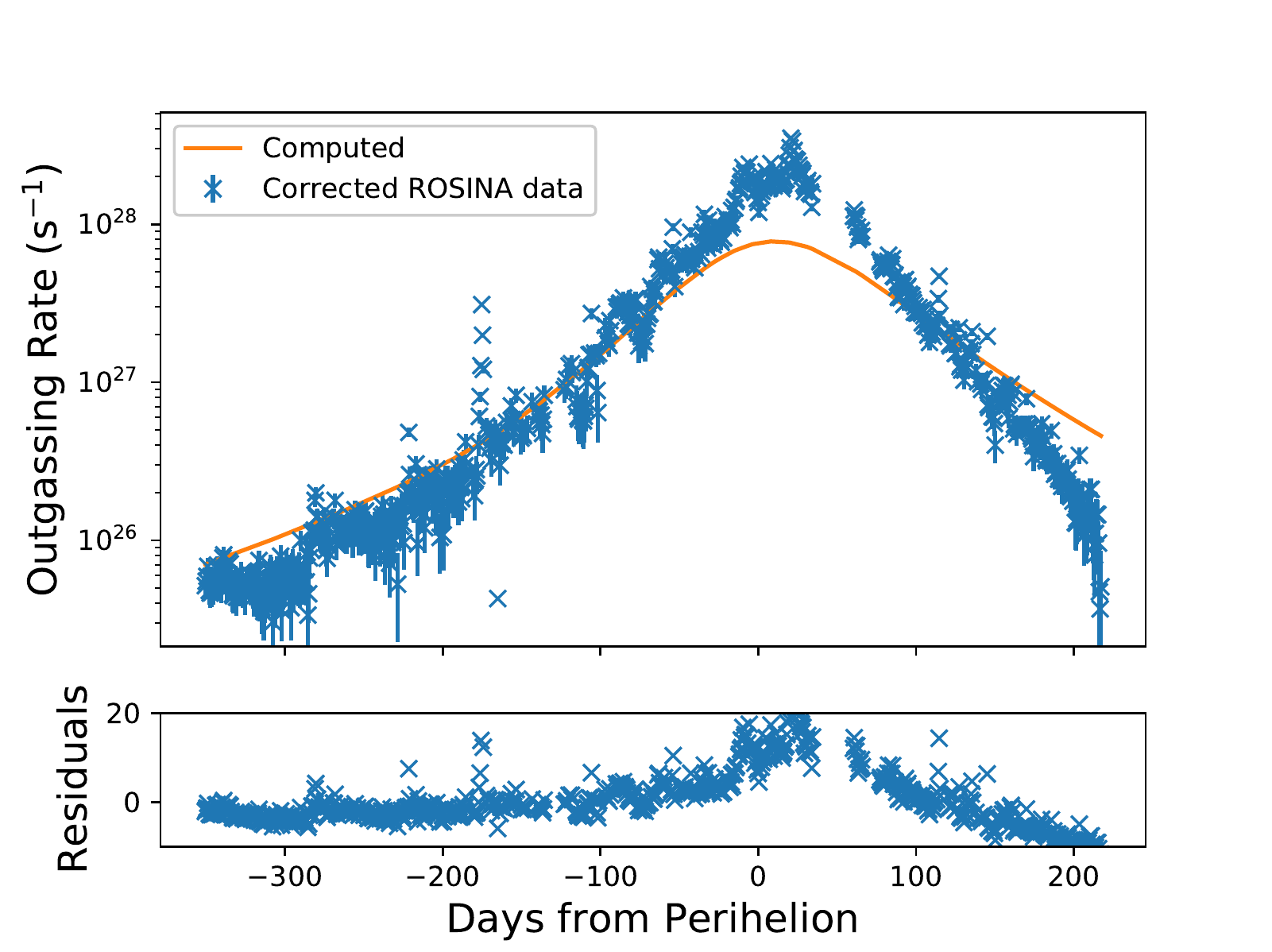}}
\caption{Observed and computed water production rates and residuals for solution A.}
\label{Q50}
\end{figure}

\begin{figure}
\resizebox{\hsize}{!}{\includegraphics{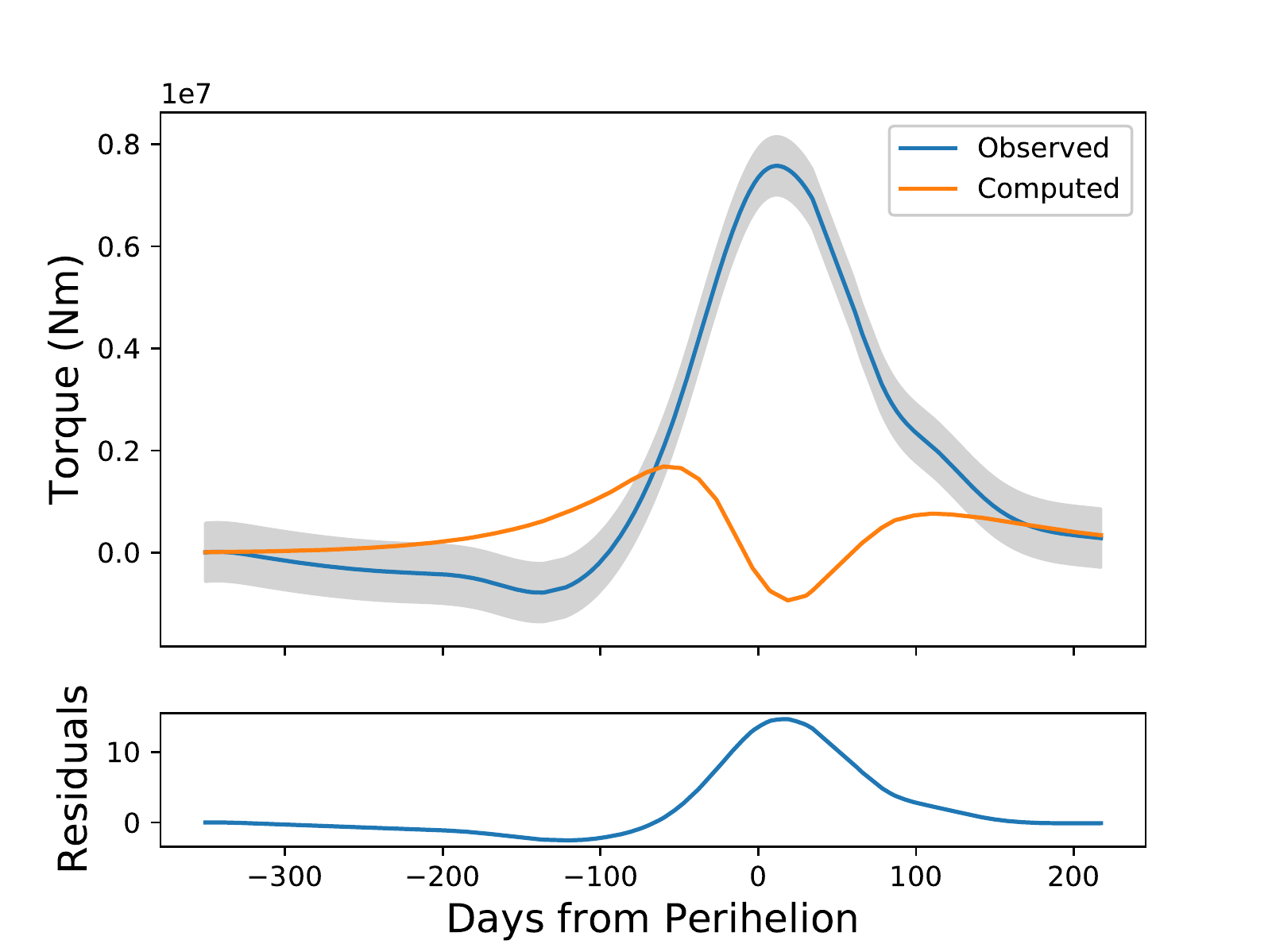}}
\caption{Observed and computed torques and residuals for solution A.}
\label{tau50}
\end{figure}

Figures \ref{Range50} --\ref{tau50} show the residuals to our best-fit solution using the five super regions defined by \cite{Marschall16, Marschall17}, which we refer to as our solution A.

\begin{figure}
\resizebox{\hsize}{!}{\includegraphics{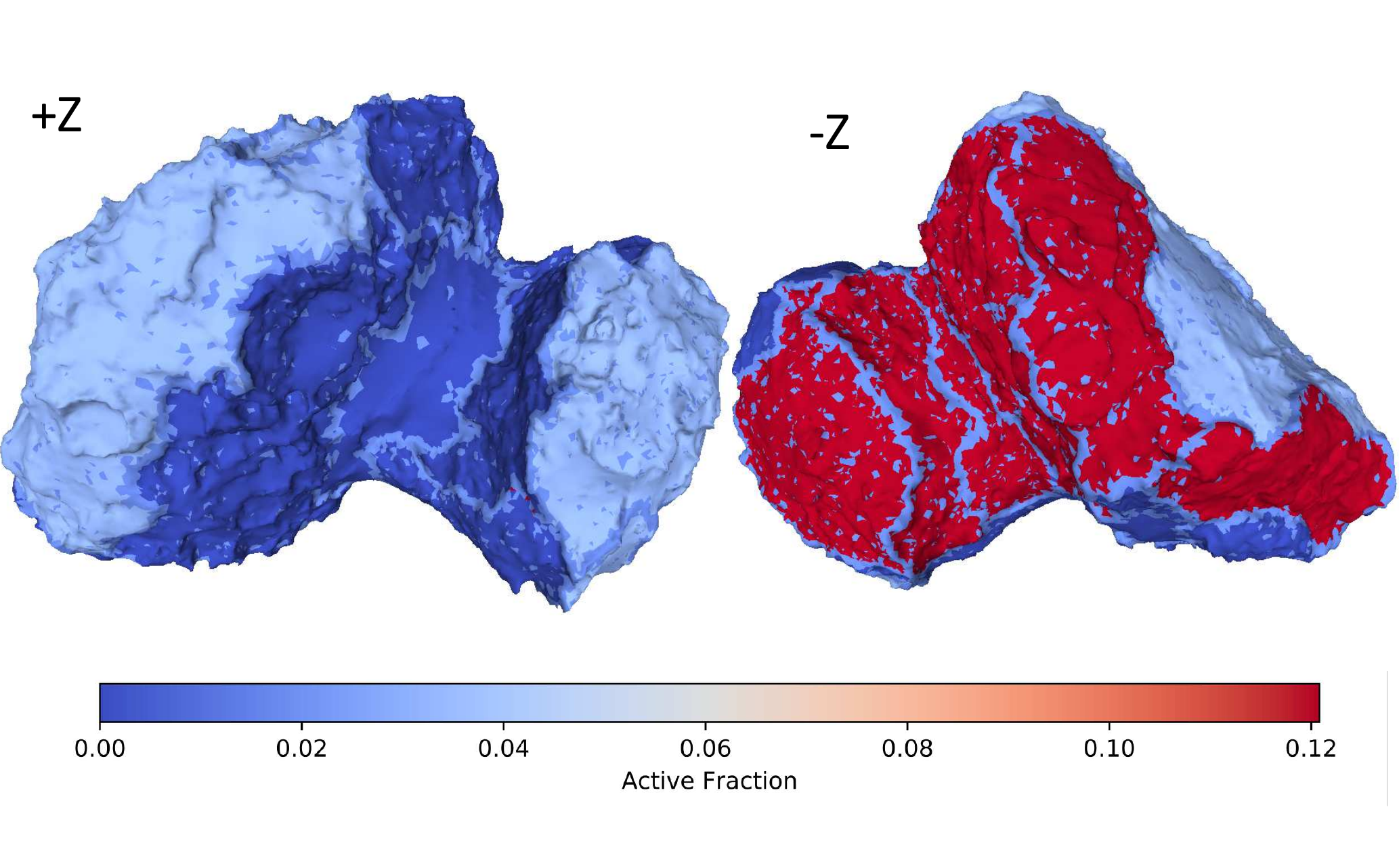}}
\caption{Mapped active fraction for solution A.}
\label{AcFrac50}
\end{figure}

The RMS range residuals, of 163 km (see Table \ref{table} for all results), represent a significant improvement over the $\sim 1000$ km of the purely gravitational solution and the $\sim 600$ km of the ground-based solution, demonstrating the significance of our NGA/N-body model. However, the water production curve is clearly not a good fit, failing to reproduce both the peak production rate as well as overestimating the production far from perihelion. Likewise, the torque curve is an extremely bad fit, failing totally to reproduce the expected positive torque peak (spin up) at perihelion. This is confirmed by the RMS (normalised) residuals of 6.59 and 5.13, respectively.

Figure \ref{AcFrac50} shows the active fractions for this solution, mapped onto the shape model. Some artefacts of the region definition can be seen, introducing spurious active fractions at the borders between regions, but these facets represent a small fraction of the total area and should not influence the general results. A large difference between the effective active fractions in the northern and southern hemispheres can be seen in Fig.~\ref{AcFrac50}, with the southern hemisphere showing active fractions of up to $12\%$, while the north is only a few percent active. This is supported by previous works \citep{Marschall16, Marschall17} based on the interpretation of ROSINA data.
It is also consistent with the higher active fraction (up to a factor of 2--3 in the southern hemisphere compared to the northern hemisphere) found by \citep{Kramer2019} from a thorough study of the evolution of the rotational parameters of 67P.
The southern regions of comet 67P receive higher insolation during southern summer which occurs near perihelion: the summer solstice happens on Aug~15, 2015, only a couple of days after perihelion.
An analysis of OSIRIS images of the Anhur/Bes southern regions \citep{Fornasier17} shows a high activity originating from these regions, combined with high relative erosion rates deduced from the relatively higher number of boulders found \citep{Pajola16}.
An examination of the production and force curves produced by the individual super regions showed that the southern hemisphere dominates production after equinox and around perihelion, as expected. Taken as a whole, the southern hemisphere has a negative torque efficiency (see Fig.~\ref{torqueefficiency}), which leads to the difficulty in jointly fitting both the production and torque curves, seen in our solution A residuals. Splitting the southern hemisphere into more regions is therefore a promising next step.

\subsection{26 parameter solution (B)}

\begin{figure}
\resizebox{\hsize}{!}{\includegraphics{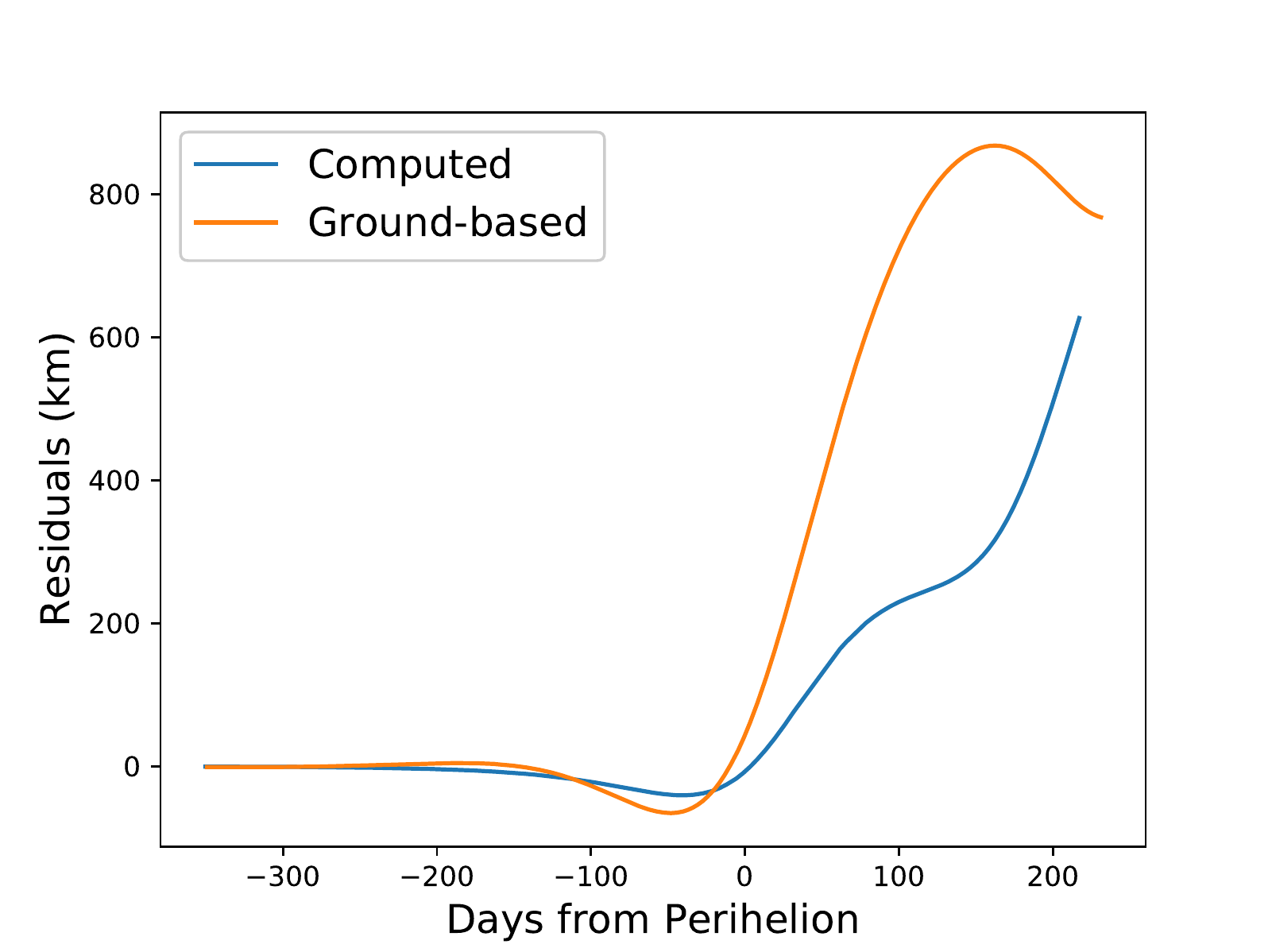}}
\caption{Observed minus computed range for solution B (blue curve), and the ground-based solution.}
\label{Range65}
\end{figure}

\begin{figure}
\resizebox{\hsize}{!}{\includegraphics{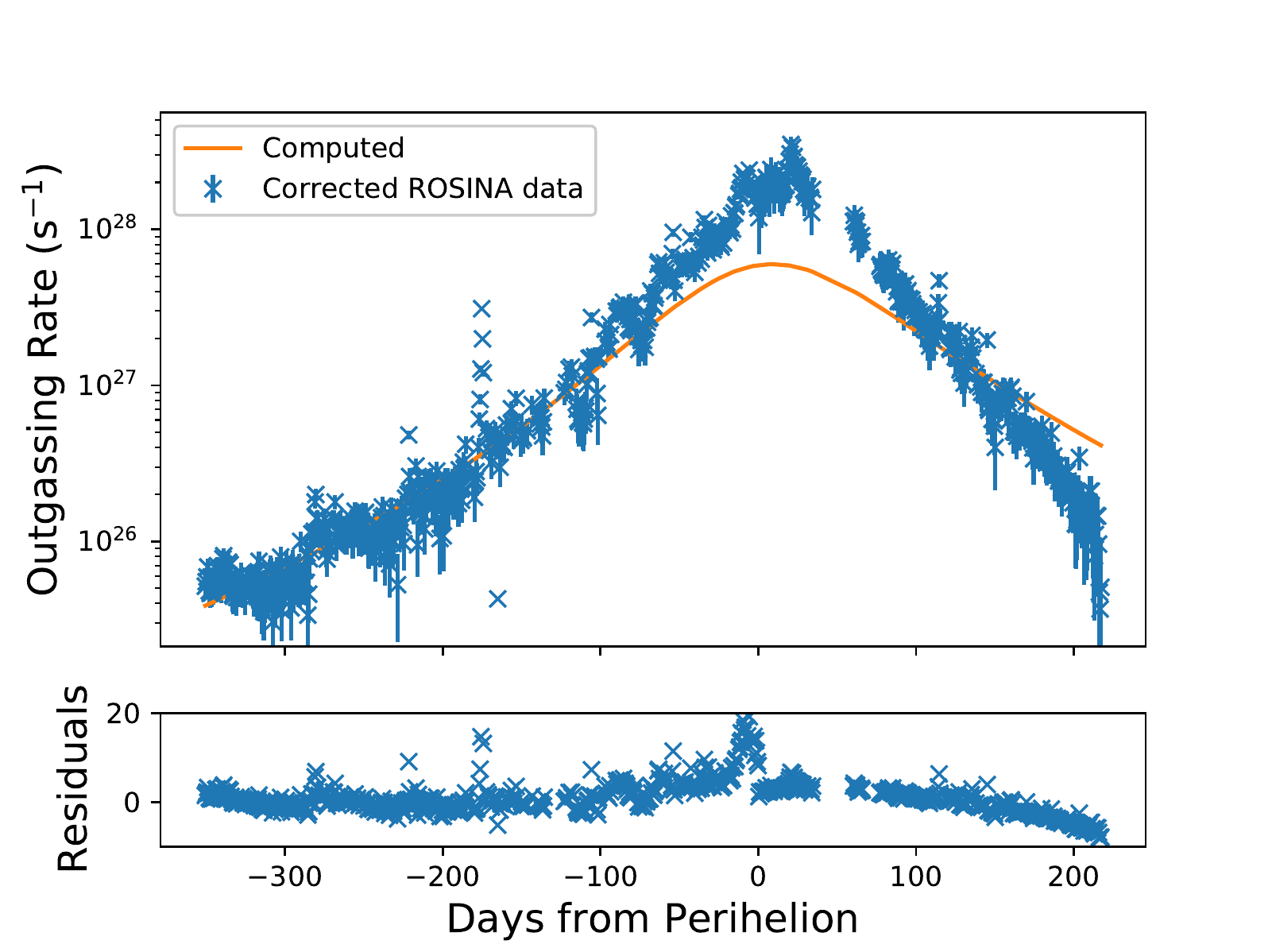}}
\caption{Observed and computed water production rates and residuals for solution B.}
\label{Q65}
\end{figure}

\begin{figure}
\resizebox{\hsize}{!}{\includegraphics{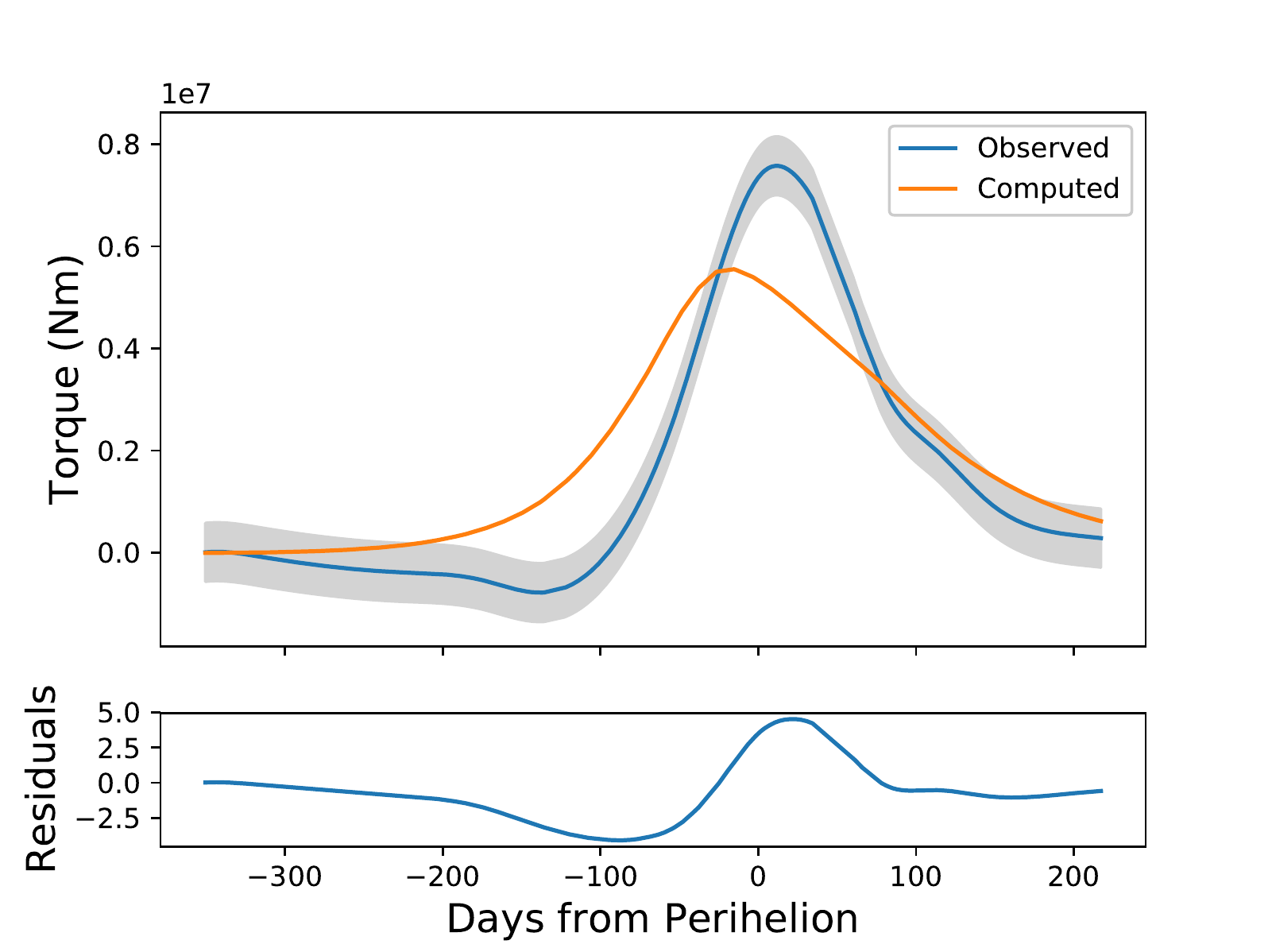}}
\caption{Observed and computed torques and residuals for solution B.}
\label{tau65}
\end{figure}

Figures \ref{Range65} --\ref{tau65} show the residuals to our optimisation with the full 26 comet regions defined by \cite{Thomas15}, which we refer to as our solution B.

\begin{figure}
\resizebox{\hsize}{!}{\includegraphics{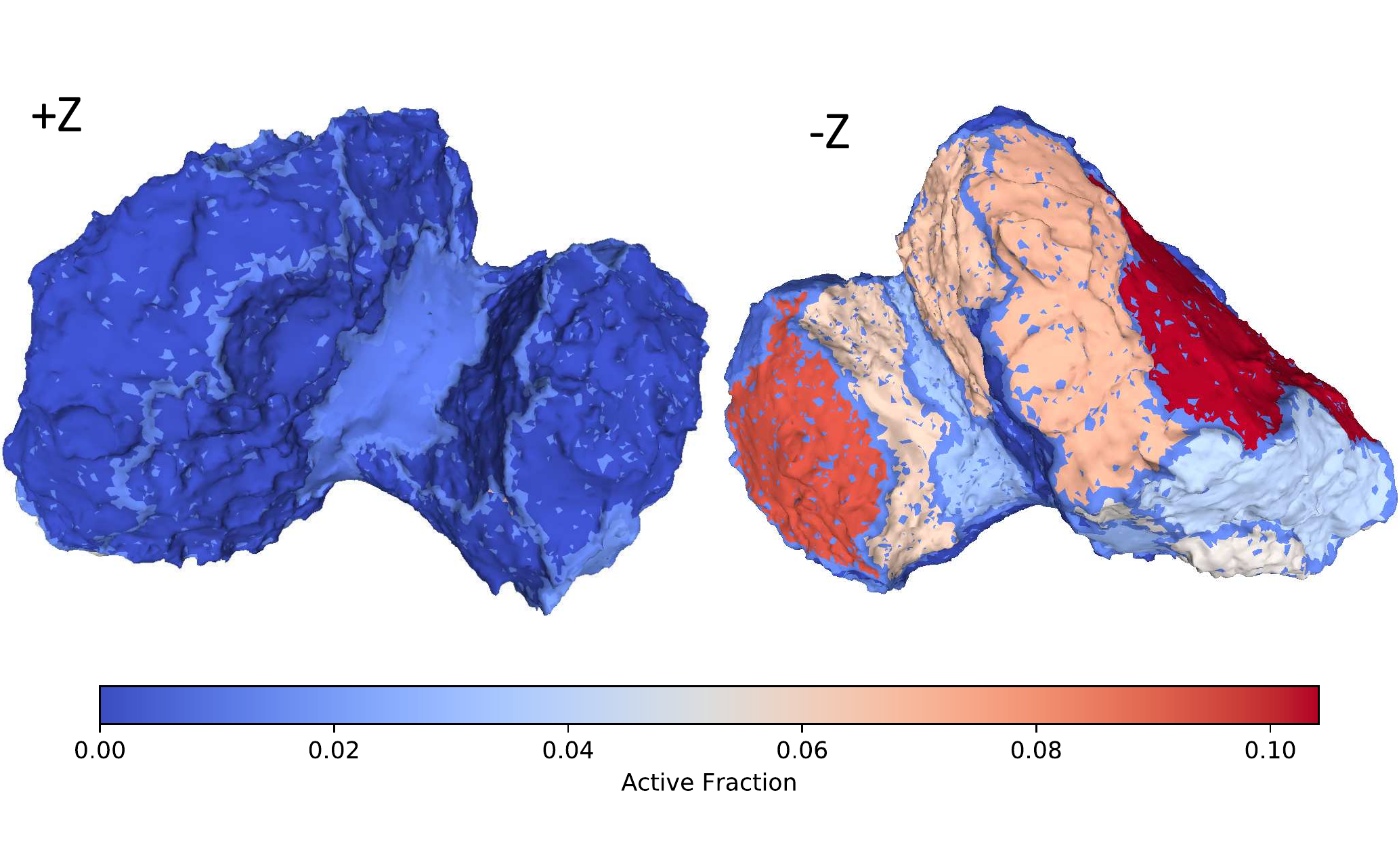}}
\caption{Mapped active fraction for solution B.}
\label{AcFrac65}
\end{figure}

Figure \ref{tau65} shows a clear improvement in how well we fit the torque peak, with the model now showing a positive peak of roughly the right magnitude, although still not matching the shape. The improved RMS residual of 2.09 backs this up. Conversely, however, the range residuals are now slightly increased, relative to solution A, and the water production curve is still not well fitted; peak production in the model is still too low, and does not fall off fast enough with heliocentric distance post-perihelion. 

The active fraction map of Fig.~\ref{AcFrac65} shows the same trend for high activity in the southern super-regions as before but now with slightly higher activity in Hapi, matching \cite{Marschall16}. Active fraction in the southern regions can be seen to vary significantly and it is instructive to compare this distribution to the map of torque efficiency. As can be seen from Fig.~\ref{torqueefficiency}, positive torque efficiency is clustered in several small areas, and these are given high activity by our optimisation. However, the optimisation is limited by the fact that the geographically defined regions containing these areas also contain areas of negative torque efficiency. In other words, torque efficiency varies at a local scale and is not necessarily correlated with the regions used in this parametrisation.

To address this, we could further subdivide our regions, introducing more parameters, but no obvious best way to do this presents itself. Instead we take the, somewhat simplified, approach of reverting back to the 5 super region model of solution A, and simply splitting the southern super region by torque efficiency. This creates two non-contiguous and ``spotty'' super regions, which do not necessarily correspond to particular morphological or structural regions on the cometary surface, but do provide a parametrised way of optimising the NGA model. Experiments with this method show significant improvements over models A and B, but still have problems reproducing the production rate curve, which we address in the next subsection. 

\subsection{Time-varying solution (C)}

Since the above solutions with constant active fractions fail to adequately reproduce all the data, we now consider time-varying solutions.
Temporal variations of the effective active fraction is an obvious way to try to reconcile the modelled post-perihelion slope of the water production rate with the measured data.

We begin with the 6 super regions (including a southern hemisphere split by torque efficiency) and implement a time-varying active fraction for both the southern hemisphere regions (since these are the most important around perihelion) while keeping the others constant. We first considered a decline of the active fraction with time, with a half-Gaussian fall-off from the initial value, fitting for both the active fractions and start and decay times of the Gaussian. While this produced encouraging results, it is a somewhat unphysical situation: the comet's orbit is cyclical so that the active fraction must ``reset'' back to the initial value in time for the next perihelion. This may happen slowly around aphelion, in which case it will not affect the fit here, or it may occur during the time-period studied by Rosetta. To explore this latter case, we perform a final optimisation, allowing two active fractions for each of the two southern hemisphere regions as well as two start times, $t_{0}$ and $t_{1}$, and two decay (or growth) half-times, $t_{\frac{1}{2}0}$ and $t_{\frac{1}{2}1}$, for a total of twelve fitted parameters. The relative magnitudes of the two active fractions are unconstrained, but the two times, $t_{0}$ and $t_{1}$, are constrained to lie either side of perihelion to ensure that they do not cross over. The half-times are constrained to be larger than zero but unconstrained at the top end. 

Finally, we also vary the momentum transfer efficiency ($\eta$ parameter) around our nominal value $\eta = 0.7$.
A full optimisation of the twelve parameters is performed adopting $\eta$ values of $0.5$, $0.6$ and $0.8$ (models Ca, Cb and Cc in Table~\ref{table}).
As shown in Table~\ref{table}, the smallest multi-objective function (Eq.~(\ref{combined_objective})) is achieved for $\eta = 0.7$, with a minimum value of $124$.
While the $\eta = 0.5$ and $\eta = 0.8$ solutions correspond to a significantly higher multi-objective function (equal to $175$ and $168$, respectively), the $\eta = 0.6$ solution (equal to $130$) is only marginally larger.

\begin{figure}
\resizebox{\hsize}{!}{\includegraphics{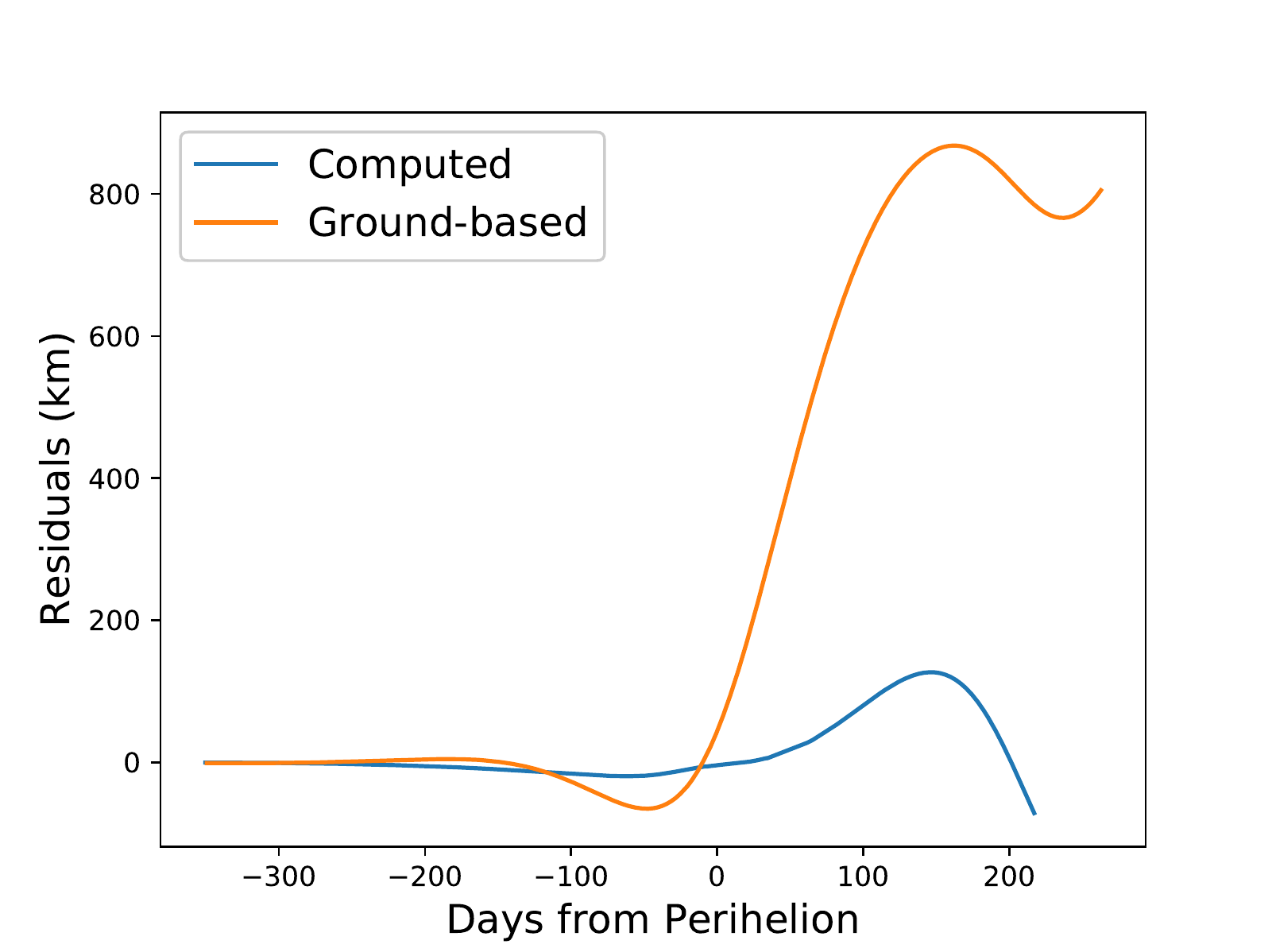}}
\caption{Observed minus computed range for solution C (blue curve), and the ground-based solution.}
\label{Range66}
\end{figure}

\begin{figure}
\resizebox{\hsize}{!}{\includegraphics{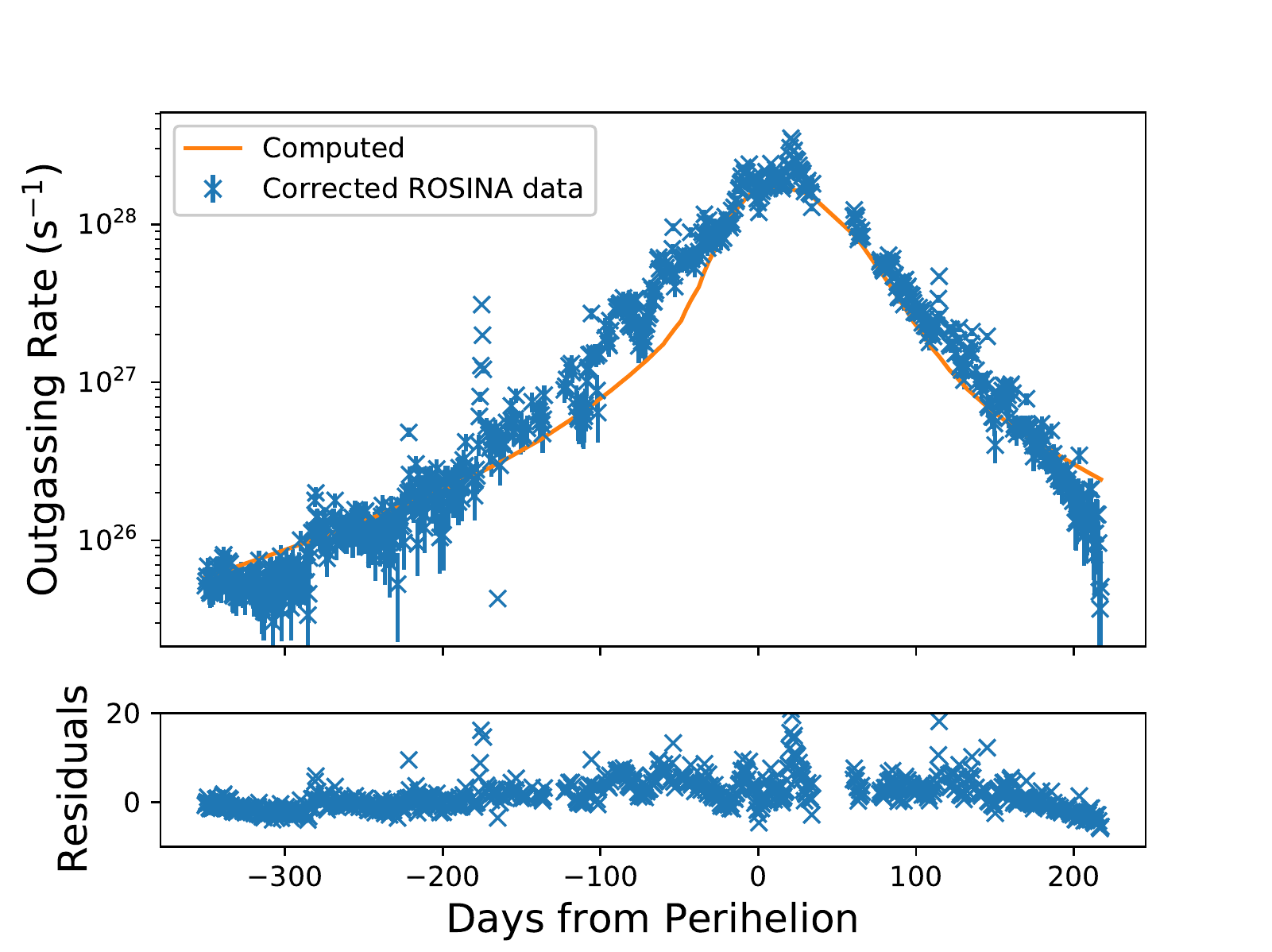}}
\caption{Observed and computed water production rates and residuals for solution C.}
\label{Q66}
\end{figure}

\begin{figure}
\resizebox{\hsize}{!}{\includegraphics{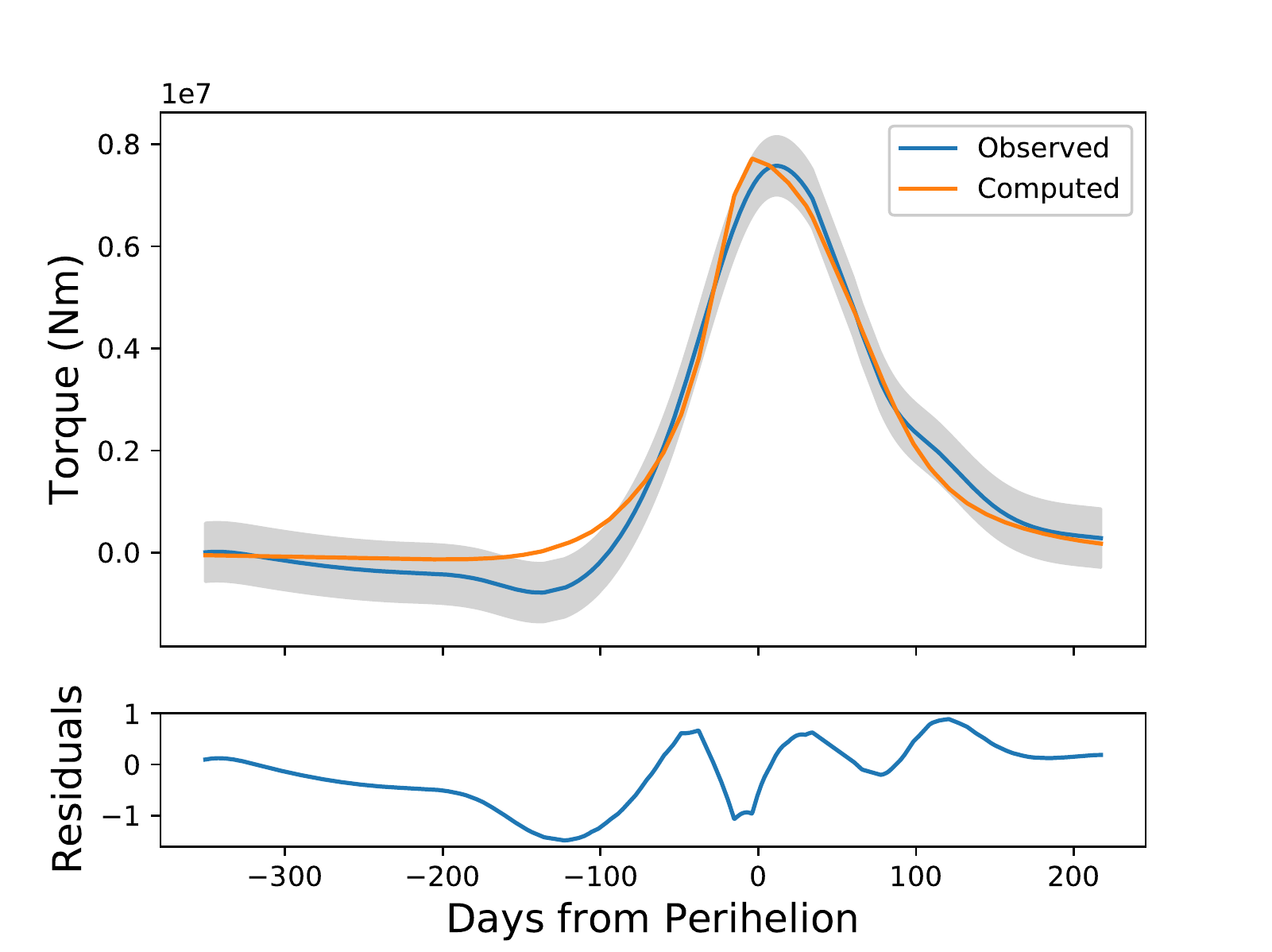}}
\caption{Observed and computed torques and residuals for solution C.}
\label{tau66}
\end{figure}

\begin{figure}
\resizebox{\hsize}{!}{\includegraphics{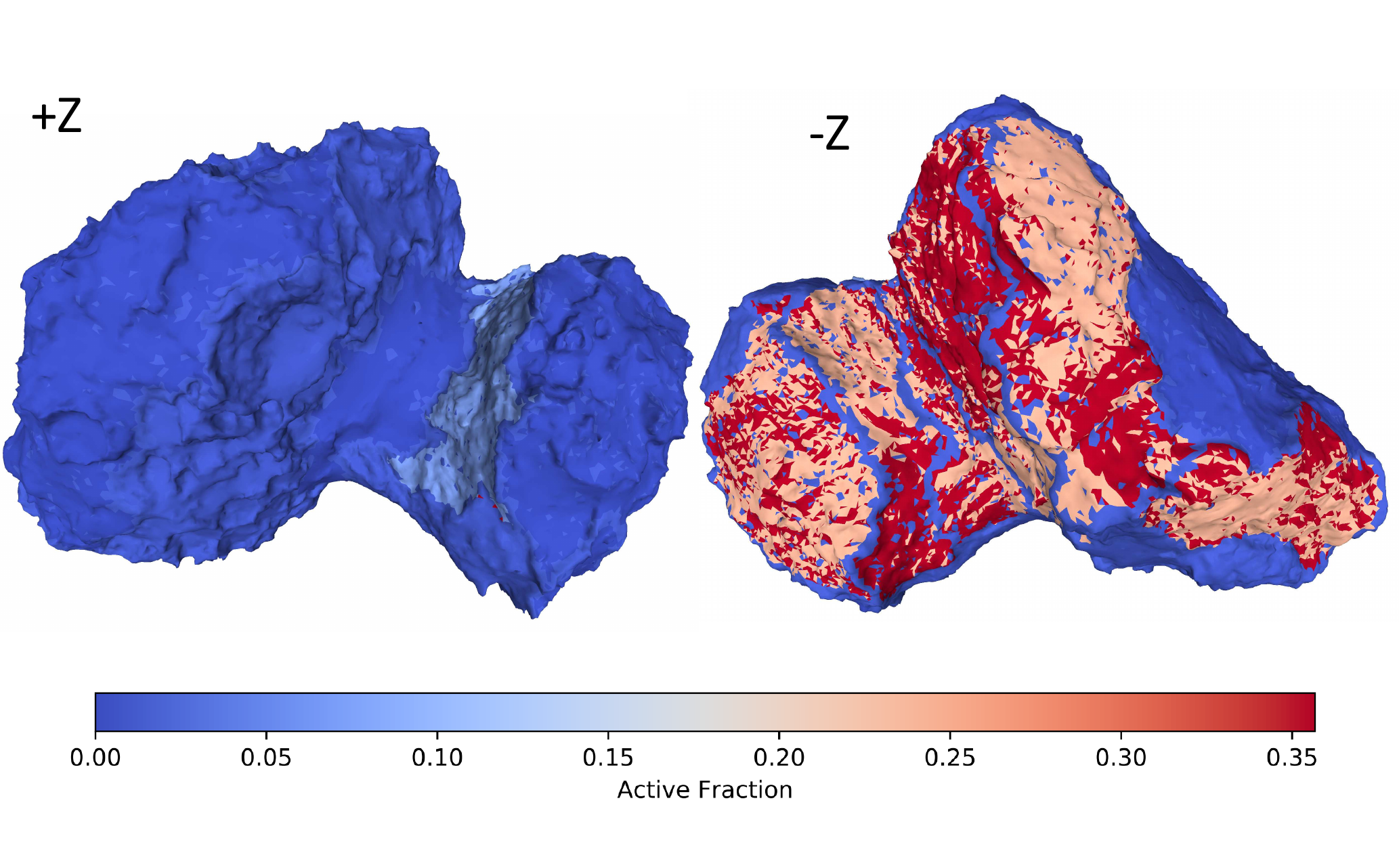}}
\caption{Mapped final active fraction for solution C.}
\label{AcFrac66}
\end{figure}

\begin{figure}
\resizebox{\hsize}{!}{\includegraphics{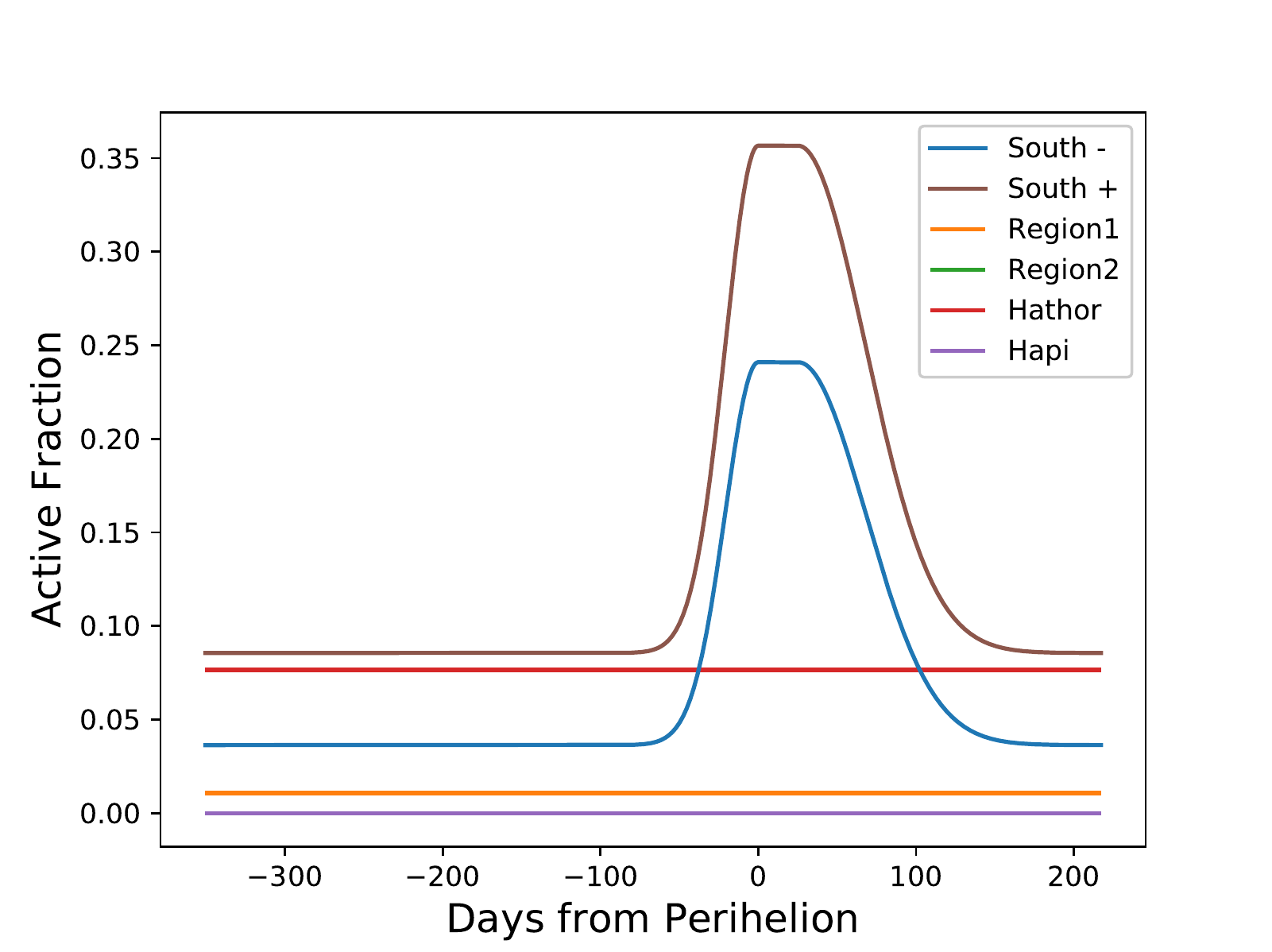}}
\caption{Active fraction with time for solution C.}
\label{AcFrac_t66}
\end{figure}

Figures \ref{Range66} --\ref{tau66} show the residuals for our best fit model in this case: model C, while Figs.~\ref{AcFrac66} and \ref{AcFrac_t66} show the mapped (peak) active fraction distribution and how it varies with time.
High activity is again favoured in the southern regions, with the optimisation now selecting very high effective active fractions, of over $35\%$, in order to fit the high peak production rates at perihelion (Fig.~\ref{Q66}).
Higher active fractions in areas producing a positive torque allow them to dominate the rest of the southern hemisphere, producing a net positive torque curve, which now matches very well the observations (Fig.~\ref{tau66}).
Note however that the small drop off observed 125~days before perihelion is not reproduced by the model.

Figure \ref{AcFrac_t66} shows that the preferred solution has southern active fractions increase quickly between equinox and perihelion (with a half-time of $\sim25$ days) to their high peak values, before falling off to the $\sim10\%$ level after perihelion and during southern summer, with a decay half-time of around 50 days. 
This reproduces the high production rate at perihelion while matching the swift fall-off over the succeeding two hundred days. Some discrepancies with the data still remain, for example the ``knee'' feature seen as the production ramps up around a hundred days before perihelion, but the result is now a much better match to the data overall.

The greatly reduced range residual, of 46~km, is further evidence of the improved model, and confirms that the majority of the NGA effect is concentrated near perihelion. The fact that non-gravitational forces and production are both seen to be strongly peaked near perihelion, over and above what one would expect from geometric considerations, is consistent with a time-varying active fraction. Small differences seen in the torque and production curves occur before equinox, when outgassing is controlled by the active fraction in the Northern hemisphere, which we do not vary with time. This suggests that minor improvements might be made to the fit by focusing on the North, although these would be unlikely to affect the trajectory and peak production, both of which are dominated by activity at perihelion (i.e.~in the South).

The best-fit results allow us to calculate integrated quantitative parameters resulting from the cometary activity around perihelion.
The total water ice mass loss amounts to $4.5\ 10^6\ \mathrm{kg}$, corresponding to a mean erosion of $\sim 9\ \mathrm{cm}$ over the entire nucleus surface, assuming a density of $538\ \mathrm{kg}\ \mathrm{m}^{-3}$ \citep{Preusker17}.
We emphasize that this estimate does not take into account the dust mass loss - predicted to be much larger depending on the dust-to-ice ratio - and the outgassing of more volatile minor species throughout the orbit.
The actual water ice erosion can be calculated by integrating the time-varying sublimation rate of each facet.
We find a local erosion of $0.4-1.4\ \mathrm{m}$ in the two southern super-regions and $< 0.1\ \mathrm{m}$ in the northern ones.

\section{Discussion}
\label{discussion}

\subsection{A temporal variation of the effective active fraction}

The most striking features of our best solution (C) are the dichotomy of the effective active fraction between the southern and northern hemispheres on the one hand, and the drastic rise of the effective active fraction around perihelion in the southern regions on the other hand.
The latter is required in our approach to explain the steep slope of the production rate.
We propose the following qualitative explanation, based on the seasonal formation and disappearance of a dust mantle, for this cyclic increase of effective active fraction in the southern regions around perihelion.
This idea has been introduced a long time ago in the literature.
Among the pioneering works, \citet{Brin1979}, \citet{Brin1980} and \citet{FanaleSalvail} introduced the idea that a dust mantle can form and be subsequently disrupted by the gas pressure if it remains thin enough.
Using a one-dimensional thermal model, \citet{Rickman90} showed how the obliquity of the spin axis can influence the stability of the mantle.
They show that a temporary mantle can form at intermediate and polar latitudes for nuclei with radii equal to $5\ \mathrm{km}$ and high obliquities (equal to $90^{\circ}$ in their simulations).
For nuclei with smaller radii (equal to $1\ \mathrm{km}$), temporary mantles only appear at perihelion distances smaller than $1\ \mathrm{a.u.}$.
In a more recent work based on a 3D thermal model, \citet{DeSanctis10} tried to reproduce the thermal evolution of 67P.
The role of the obliquity is emphasized as being critical, high obliquities favouring the appearance of a stable dust mantle at equatorial latitudes.
Other models \citep[e.g.,][]{Kossaki06} find on the contrary that a stable mantle with a non-uniform dust layer can explain the observed water production rates.

In our explanation, the southern regions - including the southern polar cap - become progressively illuminated and the activity starts to increase after the spring equinox ($93$~days before perihelion -- 11 May 2015).
This rise of activity allows dust present at the surface to be lifted off, decreasing the depth at which water ice is present below the dust layer.
This is a runaway process as the increased gas flux resulting from this reduced dust depth is able to lift off larger and larger dust particles from the surface.
An increasing fraction of these particles eventually reaches velocities large enough to escape the nucleus gravity, or to be redeposited in other nucleus (Northern) regions.
This mechanism leads to an increase of the effective active fraction.
After the summer equinox ($3$~days after perihelion -- 15 August 2015), the energy input starts to decrease, resulting in a reduced gas flux and surface temperature.
The large dust particles can no longer be lifted off from the surface and start to be redeposited locally.
The apparition of a dust mantle quenches the cometary activity as the water is no longer at the surface of the comet; instead, it sublimates through a deeper dust layer beneath the surface. 
The gas diffusion through this dust mantle produces lower gas fluxes, decreasing the effective active fraction of the surface.
The process continues for several months until the autumn equinox ($224$~days after perihelion -- 24 March 2016) triggers the southern autumn, followed by the long southern winter around aphelion.
At that time, the southern regions are covered again by a dust layer that will be partially removed at the next perihelion passage.
It is not totally excluded also that the outgassing of more volatile species during the northern summer at aphelion contributes to the re-accumulation of material in the southern hemisphere.

This scenario is partially supported by observations from instruments aboard Rosetta.
\citet{Lai16} modelled the dynamics of dust grains around 67P taking into account OSIRIS observations.
They predict that dust particles ejected from the southern hemisphere are redeposited in
the northern hemisphere during the southern summer.
A recent geomorphological analysis of the surface \citep{Birch17} also supports the hypothesis of dust being ejected from the southern hemisphere and redeposited in the northern regions, as do several other works such as \citet{Thomas15, Keller2017} and \citet{Hu17}.
Finally, \citet{Fornasier16} observed seasonal variations of the colour of the nucleus, which becomes bluer near perihelion.
This colour change is attributed to the removal of the dust mantle covering regions with low gravitational slopes as the comet approaches perihelion.

It must be noted that our model interprets active area as a literal fraction of the surface area which is covered with water ice and is outgassing. Since exposed water ice is rare on the surface, this is a simplification of the real process, which should involve gas flow through pores, dust layers, erosion etc., the details of which we consider beyond the scope of this paper. Nonetheless, effective active fraction is a useful proxy with which to quantify activity, with the above-noted spatial and temporal variations providing an insight into the changing activity patterns over the comet's seasons. Our area-weighted global average values, of $6.4\%$ around perihelion and $1.9\%$ otherwise, agree well with previous estimates for 67P \citep{Lamy2007} and other comets (see, e.g.~\citealp{AHearn1995}), but the large differences between hemispheres and seasons highlights the limitations of interpreting cometary activity with a single number from the ground.

\subsection{A refined value for the $\eta$ parameter}

The $\eta$ parameter, also called ``momentum transfer efficiency'' in the literature represents the fraction of the Maxwellian thermal velocity of the gas, calculated from the nucleus surface temperature, contributing to the momentum transfer.
This parameter depends on the local ice content and on the detailed structure of the porous material.
Its value couples the water production curve with the effect of the spin/orbit variations, allowing to compensate any systematic effect possibly present in the input data or in the model. 
The value of the $\eta$ parameter has been calculated from gas-kinetic models of the Knudsen layer.
\citet{Delsemme71} adopted a value $\eta = 0.6$, between the pure ice plane surface value $\eta = 1/2$ and higher values up to $2/3$ predicted for porous media.
In early interpretations of the NGA of comet~1P/Halley, \citet{Rickman86} and \citet{Sagdeev88} used a value of $0.5$ and $0.79$ respectively based on different assumptions.
\citet{Crifo1987} recommended a value $\eta \approx 0.5$ based on revised gas-kinetic theoretical description of the solid-gas interface, taking into account the recondensation of water ice.
\citet{Rickman89} used a corrected value $\eta = 0.53-0.67$ based on the work of \citet{Crifo1987}, which seem to be in good agreement with gas velocity measurements in the coma of comet 1P/Halley \citep[see][and references therein]{Rickman89}.
However, the calculations do not consider intimate ice-dust mixtures and do not take into account the porosity of the surface.
\citet{Skorov99} introduced a correction factor of $1.8$ to the values adopted by \citet{Rickman89}
based on an analytical model of the Knudsen layer above a porous dust mantle.
This leads to very high $\eta$ values in the range $1-1.2$.

From our analysis of the data presented in this paper, our best fits are obtained for $\eta$ in the range $0.6-0.7$, in good agreement with the moderate values adopted in the literature.
They do not support more extreme values (around $0.5$ and greater than $0.8$), which degrade the overall fit of the three data sets.
We stress however that we set the surface gas temperature to the dust temperature $T_{dust}$ in our model (see Eq.~(\ref{v})).
This may lead to overestimating the gas temperature, which would in turn tend to underestimate the fitted 
value of $\eta$.
Note, however, that the dependence of the gas velocity with the temperature is a square root in Eq.~(\ref{v}): a large and constant
temperature deviation would be needed to significantly bias the value of $\eta$.
The second point of concern is a possible overestimate of the water production curve due to sublimating icy grains in the coma.
This would once again require a larger value of $\eta$ to compensate the smaller local sublimation rates.
Altogether, even if our simulations point to $\eta < 0.8$ we cannot entirely rule out at this point higher values of
this parameter.
Consolidated values of the surface water production around perihelion, as well as estimates of the gas velocity above the
surface, would help to reduce the uncertainties.

\section{Conclusions and perspectives}

From our work, the following conclusions could be reached:
\begin{enumerate}
\item We succeeded in finding an activity pattern explaining simultaneously the three following 67P data sets, extracted mainly from the Rosetta mission: (1)~the Earth-comet ranging data reconstructed by the flight dynamics team of ESA/ESOC, (2)~the water production rate deduced from a mix of ROSINA and ground-based observations \citep{Hansen}, and (3)~the rate of spin period change deduced from the OSIRIS images.
The residuals of the ranging data describing the effect of the non-gravitational acceleration are reduced by an order of magnitude compared to the ground-based solution based on the simple model of \citet{Marsden1973}.
\item The fitted activity pattern exhibits two main features: a higher effective active fraction in two southern super-regions ($\sim 10$~\%) outside perihelion compared to the northern ones ($< 4$~\%), and a drastic rise of the effective active fraction of the southern regions ($\sim 25-35$~\%) around perihelion.
\item In order to successfully fit the positive rate of spin period change, we need to split the southern super region into two entities, depending on the sign of the torque efficiency.
These two entities correspond to two relatively well delimited areas in Anhur, Bes and Khepry, but creates a patchy separation in Wosret.
\item We interpret the time-varying southern effective active fractions by cyclic formation and removal of a dust mantle in these regions.
According to our interpretation, the dust mantle could be progressively removed when activity rises after the southern spring equinox and formed again when activity decreases towards the southern autumn equinox (and possibly around aphelion during the northern summer).
Several observations performed during the Rosetta mission, such as dust transport from South to North \citep{Lai16,Birch17} and bluer colours observed near perihelion \citep{Fornasier16}, support this interpretation.
\item If it is confirmed, this interpretation would strongly support post-Halley thermal modelling \citep{Rickman90,DeSanctis10} which predicted that seasonal effects linked to the orientation of the spin axis play a key role in the formation and evolution of dust mantles, and in turn largely control the temporal variations of the gas flux.
\item Our analysis supports moderate values of the momentum transfer coefficient $\eta$ in the range $0.6-0.7$.
For more extreme values of this coefficient ($\leq 0.5$ and $\geq 0.8$), the fit of the three data sets is degraded.
However, we will not be able to rule out higher values of this parameter without consolidated water production measurements and,
to a lesser extent, without estimates of the near-surface thermal gas velocity.
\end{enumerate}

More work will be needed to better understand the activity of 67P using data collected during the Rosetta mission.
The solution found in this article through the process described in section~\ref{results} is non-unique on the one hand, and could probably be improved on the other hand.
Improvements may come from a better understanding of the ranging data, resulting in the extraction of clean and accurate non-gravitational acceleration of the comet as a function of time around perihelion.
The change in the direction of the spin axis or angular velocity could also provide a valuable additional data set that could help to constrain the activity pattern and reduce the number of solutions. 
In their interpretation of the temporal evolution of the rotational parameters, \citet{Kramer2019} did not introduce a temporal variation of the activity around perihelion but considered instead a spatially heterogeneous surface with 36 ``patches'' having different water-ice coverage.
They also discuss the possibility of a decreasing dust layer near perihelion increasing activity, and found a solution explaining the water production curve of 67P.
It would be interesting to test if this solution could also reproduce the NGA of comet 67P described in this article.
Finally, a better understanding of the production rate around perihelion, reconciling the different Rosetta instrument measurements and including species other than water, would be of benefit in fitting the activity model.

\begin{acknowledgements}
This project has received funding from the European Union's Horizon 2020 research and innovation programme under grant agreement no. 686709. This work was supported by the Swiss State Secretariat for Education, Research and Innovation (SERI) under contract number 16.0008-2. The opinions expressed and arguments employed herein do not necessarily reflect the official view of the Swiss Government. We thank the authors of several python package libraries which were used extensively here; these include \textit{NumPy, SciPy, SpiceyPy, REBOUND} and \textit{REBOUNDx}. We are grateful to L. Maquet for fruitful discussions in the course of this work. We also thank the reviewer, Dennis Bodewits, for his comprehensive and useful review.
\end{acknowledgements}

\bibliographystyle{aa}
\bibliography{Bibliography}

\end{document}